\documentclass[journal=jacsat,manuscript=article]{achemso}

\usepackage[version=3]{mhchem} % Formula subscripts using \ce{}

\usepackage{xcolor}
\usepackage{soul}
\usepackage[colorinlistoftodos]{todonotes}
\author{Sk Mujaffar Hossain}
\affiliation[IKST]
{Indo-Korea Science and Technology Center (IKST), Bangalore 560064, India}

\author{Namitha Anna Koshi}
\affiliation[IKST]
{Indo-Korea Science and Technology Center (IKST), Bangalore 560064, India}

\author{Seung-Cheol Lee}
\email{leesc@kist.re.kr}
\affiliation[IKST]
{Indo-Korea Science and Technology Center (IKST), Bangalore 560064, India}
\alsoaffiliation[EMRC KIST]
{Electronic Materials Research Center, Korea Institute of Science $\&$ Technology, Seoul 136-791, South Korea}

\author{G.P Das}
\email{gour.das@tcgcrest.org}
\affiliation[RISE TGC CREST]
{Research Institute for Sustainable Energy (RISE), TCG-CREST, Kolkata 700091, India}

\author{Satadeep Bhattacharjee}
\email{s.bhattacharjee@ikst.res.in}
\affiliation[IKST]
{Indo-Korea Science and Technology Center (IKST), Bangalore 560064, India}
\title[An \textsf{achemso} demo]
  {Integrating Density Functional Theory with Deep Neural Networks for Accurate Voltage Prediction in Alkali-Metal-Ion Battery Materials}%\footnote{A footnote for the title}}

\abbreviations{IR,NMR,UV}

\begin{document}

%\begin{tocentry}

%Some journals require a graphical entry for the Table of Contents.
%This should be laid out ``print ready'' so that the sizing of the

%Inside the \texttt{tocentry} environment, the font used is Helvetica
%8\,pt, as required by \emph{Journal of the American Chemical

%The surrounding frame is 9\,cm by 3.5\,cm, which is the maximum
%permitted for  \emph{Journal of the American Chemical Society}
%graphical table of content entries. The box will not resize if the
%content is too big: instead it will overflow the edge of the box.

%This box and the associated title will always be printed on a
%separate page at the end of the document.

%\end{tocentry}

\begin{abstract}
Accurate prediction of the voltage of battery materials plays a pivotal role in the advancement of energy storage technologies and the rational design of high-performance cathode materials. In this work, we present a deep neural network (DNN) model, built using PyTorch, to estimate the average voltage of cathode materials across Li-ion, Na-ion, and other alkali-metal-ion batteries. The model is trained on an extensive dataset from the Materials Project, incorporating a wide range of specific structural, physical, chemical, electronic, thermodynamic, and battery descriptors, ensuring a comprehensive representation of material properties. Our model exhibits strong predictive performance, as corroborated by first-principles density functional theory (DFT) calculations. The close alignment between the DNN predictions and the DFT outcomes highlights the robustness and accuracy of our machine learning framework to effectively select and identify viable battery materials. Using this validated model, we successfully proposed novel Na-ion battery compositions, with their predicted behavior confirmed by rigorous computational assessment. By seamlessly integrating data-driven prediction with first-principles validation, this study presents an effective framework that significantly accelerates the discovery and optimization of advanced battery materials, contributing to the development of more reliable and efficient energy storage technologies.

 %Accurately predicting the voltage of battery materials is essential for advancing energy storage technologies and designing more efficient, high-performance batteries. In this study, we developed a deep neural network (DNN) model to predict the average voltage of materials used in Li-ion, Na-ion, and other alkali-metal-ion batteries. A comprehensive dataset was compiled from the Materials Project, incorporating a diverse set of features, including structural, physical, chemical, electronic, and thermodynamic properties, along with battery-specific descriptors. These features were utilized to construct a robust DNN model aimed at facilitating the discovery of novel battery materials. The model performance evaluated through 10-fold cross validation, achieving an  R$^2$ value of 0.99 and a mean absolute error (MAE) of 0.069 V on the validation dataset. This high level of accuracy underscores the model's capability to capture the complex relationship between material properties and electrochemical performance. Furthermore, the results demonstrate that machine learning approaches, particularly DNN models, can provide rapid and reliable voltage predictions without relying on computationally expensive first-principles calculations.
%This research shows how systematic data analysis can accelerate the identification and improvement of next-generation battery materials. By focusing on key material properties and using computational methods, our study lays the groundwork for faster development of new energy storage solutions.
\end{abstract}

\section{Introduction}
In different disciplines of science and technology, the rational design of modern materials is the ultimate goal. During the past one to two decades, the materials science community has made a tremendous effort to compile the large data set of materials with different properties from multiple sources\cite{andersen2021optimade,evans2024developments,evans2021optimade}  so as to provide easy access to the database repository for scientists and engineers to aid in the discovery of novel materials using artificial intelligence (AI) and machine learning (ML). In every domain such as material science, medical science, social science, advanced technology, etc., there is a growing demand for new high-performance, highly efficient, and robust materials. %To meet this demand, researchers have intensified their efforts in various fields of material science, such as high temperature superconductivity\cite{gashmard2024predicting,seegmiller2023discovering,yazdani2023artificial}, solar cell \cite{hui2023machine,bansal2023machine,bhatti2023revolutionizing,valsalakumar2024machine}, energy storage and beyond. 
In recent decades, the Li-ion battery (LIB) has been used as a prominent renewable energy storage device and is dominant in all major electronic applications, from tiny watch to large battery in the EV sector and captures most of the shares in the commercial market\cite{ren2023comprehensive}. However, different environmentally friendly renewable energy sources (solar, wind, etc.) and their utilization demand the need for higher density and highly durable energy storage materials. The high manufacturing cost, low abundance of Li metal and the paucity of other constituent metallic elements used in the electrode, and other factors hinder the further commercialization of existing battery technology to meet the current demand\cite{ren2023comprehensive}. To overcome these challenges, researchers are moving towards anode-free metal ion batteries\cite{hatzell2023anode,godbole2024light} and other earth abundant active metal ion batteries such as Na, Zn, etc.\cite{hu2023emerging,thirupathi2023recent,zhao2023structure,nie2023recent}. 

Developing new electrode materials with high energy density and long cycle life, comparable to lithium-ion batteries (LIBs), remains a significant challenge. The process is not only technically demanding but also costly and time consuming due to the complexities of laboratory synthesis and experimental validation. To solve these issues, researchers are trying to use fast-growing and high-demand AI\cite{pandey2023artificial} and ML\cite{sarker2021machine} techniques with the latest updated battery data repository such as Materials Project (MP)\cite{jain2013commentary}, OQMD\cite{kirklin2015open,saal2013materials}, AFlowLib\cite{curtarolo2012aflowlib}, ESP\cite{ortiz2009data}, CMR\cite{landis2012computational}, NOMAD\cite{draxl2018nomad}, ICSD\cite{bergerhoff1983inorganic}, COD\cite{gravzulis2012crystallography}, NASA\cite{bole2014adaptation,hogge2015verification}, etc. 

Machine learning (ML) has emerged as a powerful tool for screening potential materials (with applications in battery, catalysis, solar cells, etc.) due to its ability to significantly accelerate the discovery process compared to traditional density functional theory (DFT) calculations\cite{taouti2025dft}. Although DFT provides high-fidelity electronic structure insights and reliable predictions of material properties, it is still computationally expensive, especially for large-scale materials exploration\cite{nagai2020completing}. In contrast, ML models, once trained on high-quality DFT or experimental data, can predict key material properties with orders of magnitude greater efficiency, allowing for the rapid screening of vast chemical spaces\cite{taouti2025dft,yin2022deep}. ML is particularly advantageous for identifying promising candidates early in the design process, reducing the need for exhaustive DFT calculations\cite{zuccarini2024material}. However, ML models are highly dependent on the quality and diversity of their training data, and their predictions may lack the fundamental physical interpretability of first-principles methods such as DFT\cite{chan2022application}. Despite this, integrating ML with DFT in a hybrid approach can enhance both accuracy and efficiency, utilizing the robustness of DFT while leveraging the speed of ML to optimize the search for high-performance battery materials.

Significant progress has recently been achieved in the search for new electrode materials, through prediction of battery voltages\cite{joshi2019machine,moses2021machine}, volume change\cite{moses2021machine}, chemical reaction\cite{louis2022accurate}, and formation energy\cite{louis2022accurate} of electrode materials using a variety of ML techniques in conjunction with the well-captured clean battery database. This database\cite{evans2021optimade, evans2024developments} has been predominantly created using quantum-mechanical DFT calculations. By combining the DFT structured database with improved and sophisticated ML algorithms, one can accelerate the discovery of modern electrode materials. 
Various regression- and classification-based ML models have been employed for both the design and prediction of new electrode materials. For example,
\citeauthor{joshi2019machine} utilized deep neural networks (DNNs), support vector regression (SVR), and kernel ridge regression (KRR) to predict electrode voltages for metal-ion batteries, achieving a mean absolute error (MAE) of approximately 0.43 V after 10-fold cross-validation\cite{joshi2019machine}, \citeauthor{mishra2025designing} employed XGBoost for the prediction of the average voltage in battery materials and reported a root mean square error (RMSE) of 0.41 V\cite{mishra2025designing}. Similarly, \citeauthor{moses2021machine} developed a regression-based DNN model considering the volume change of electrode materials during the charge and discharge cycles, reporting an MAE of 0.47 V for average voltage predictions \cite{moses2021machine}. Moreover, \citeauthor{louis2022accurate} employed Graph Neural Networks (GNNs) to predict battery voltages using two different strategies, chemical reaction energy and formation energy-based approaches, obtaining MAE values in the range of 0.31 to 0.34 V\cite{louis2022accurate}. 

Beyond voltage prediction, ML-driven research has also expanded toward the rational design of high-energy-density cathode and anode materials, where models are trained to screen novel compositions with optimized electrochemical stability and fast ion diffusion kinetics\cite{sendek2017holistic}. In addition, ML techniques are being leveraged for the discovery of solid-state electrolytes (SSEs), with the aim of improving ionic conductivity while maintaining chemical and electrochemical stability \cite{sendek2017holistic, li2024machine, mishra2023exploring}. Liquid electrolytes have also been explored using ML approaches\cite{hu2023machine}, particularly in predicting solvation energy \cite{ferraz2023explainable,alibakhshi2021improved, ward2021graph}, electrochemical window\cite{manna2023molecular,zhang2014refined}, and ionic mobility\cite{baskin2022benchmarking,li2024machine}, which are crucial to improve battery safety and performance \cite{hui2023machine, hu2023machine}. Another important avenue is battery health diagnostics and lifetime prediction\cite{thelen2024probabilistic,zou2024machine,sekhar2023prediction}, where ML-based models analyze degradation mechanisms, capacity fade trends, and structural stability over prolonged cycling. As AI and ML techniques continue to evolve with larger datasets, improved feature selection strategies, and more sophisticated deep learning architectures, their role in battery material discovery and optimization will only grow stronger. The combination of DFT-driven data generation and ML-guided material screening has already proven to be a powerful synergy, paving the way for the next generation of high performance, durable, and sustainable energy storage materials\cite{ling2022review}.

The development of high-energy-density Li-ion and Na-ion battery materials has gained significant attention in both academia and industry.  However, complex phase transitions during insertion and extraction pose challenges, such as poor cycle stability, low energy density, and limited rate capability. To address these challenges, researchers have explored various strategies for optimizing Li/Na-ion cathodes. In this work, we leverage a DNN model developed using the PyTorch library to design novel layered transition-metal oxide-based compositions. Our model is robust and highly predictive, outperforming previously published approaches for voltage prediction\cite{louis2022accurate,joshi2019machine,moses2021machine}. Feature engineering plays a crucial role in materials informatics, where properties derived from chemical, structural, elemental, electronic, and thermodynamic representations are used to construct feature vectors. By incorporating these effective strategies, we trained our DNN model to predict the voltage of electrode materials with high accuracy. As a result, we identified several promising cathode materials for Na-ion batteries with high voltage and energy density. To validate our predictions, we further conducted first-principles DFT simulations, which confirm the potential of these materials for next-generation Na-ion batteries.

\section{The Rise of Na-Ion Batteries in Energy Storage}
\begin{figure}[htp]
    \centering
\includegraphics[width=15cm]{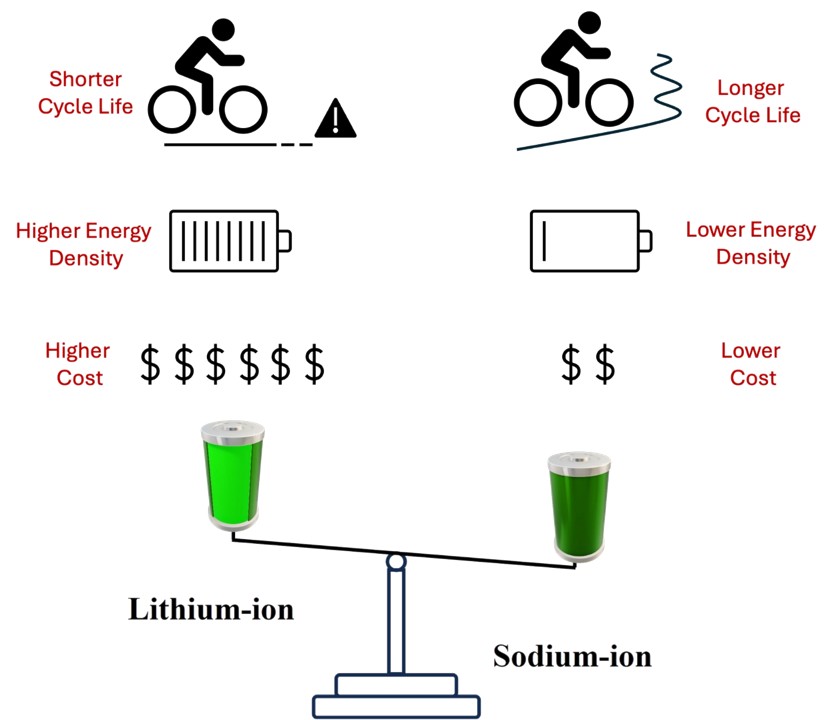}
    \caption{Schematic illustration highlighting the differences between Li-ion and Na-ion batteries.}
    \label{fig:fig_01}
\end{figure}

Researchers are increasingly focusing on sodium-ion batteries (Na-ion) as a promising alternative to lithium-ion (Li-ion) technology due to several key advantages in cost, resource availability, and sustainability (Figure \ref{fig:fig_01}). One of the most significant benefits of Na-ion batteries is their lower material cost, with a theoretical cost of \$40-77/kWh, making them 30\% cheaper than LiFePO$_4$ based lithium-ion batteries\cite{tarascon2020ion,abraham2020comparable}. This cost advantage is particularly crucial given the increasing costs and scarcity of lithium, which has led to an increase in production expenses for Li-ion batteries, reaching an average of \$137/kWh in 2020. In terms of energy density, Na-ion batteries offer up to 290 Wh/kg for active materials and 250-375 Wh/L volumetric energy density, making them competitive for energy storage applications. However, current practical Na-ion prototypes have a lower energy density (75-200 Wh/kg) compared to lithium-ion equivalents, but continuous research and development are expected to improve these values.

Another critical factor driving the research of Na-ion batteries is the cycle life. Although early prototypes have shown faster degradation, some advanced Na-ion battery designs have demonstrated 4,000 cycles, rivaling Li-ion batteries\cite{tarascon2020ion,chayambuka2020li}. This long cycle life is particularly attractive for applications such as grid storage, where energy density is less critical than longevity and cost-effectiveness. In general, the growing interest in Na-ion batteries stems from their potential to provide a sustainable, cost-effective, and scalable energy storage solution without the resource limitations associated with lithium. As research progresses, improvements in energy density, stability, and charge-discharge performance could make Na-ion batteries a viable alternative to Li-ion batteries for a wide range of applications, including renewable energy storage, electric vehicles, and large-scale power grids. As a result, the development of Na-ion battery materials has gained significant attention in both academia and industry. 

Among various Na-ion cathode materials, many industries (such as HiNa, CATL, Faradian, Tiamat, etc.) have focused on layered transition metal oxides of O3 and P2-types, as they emerged as highly promising due to their superior capacity compared to polyanionic and Prussian blue-based cathodes\cite{sayahpour2022perspective}. However, the large ionic size of Na$^+$ slows kinetics and induces complex phase transitions during insertion and extraction, causing challenges such as poor cycle stability, low energy density, and limited rate capability. To address these challenges, researchers have explored various strategies for optimizing Na-ion cathodes. In this work, we leverage a DNN model developed using the PyTorch library to design novel O3- and P2-type layered transition metal oxide compositions. 

\section{Methods}

\subsection{Dataset collection and descriptors}
\begin{figure}[htp]
    \centering
\includegraphics[width=15cm]{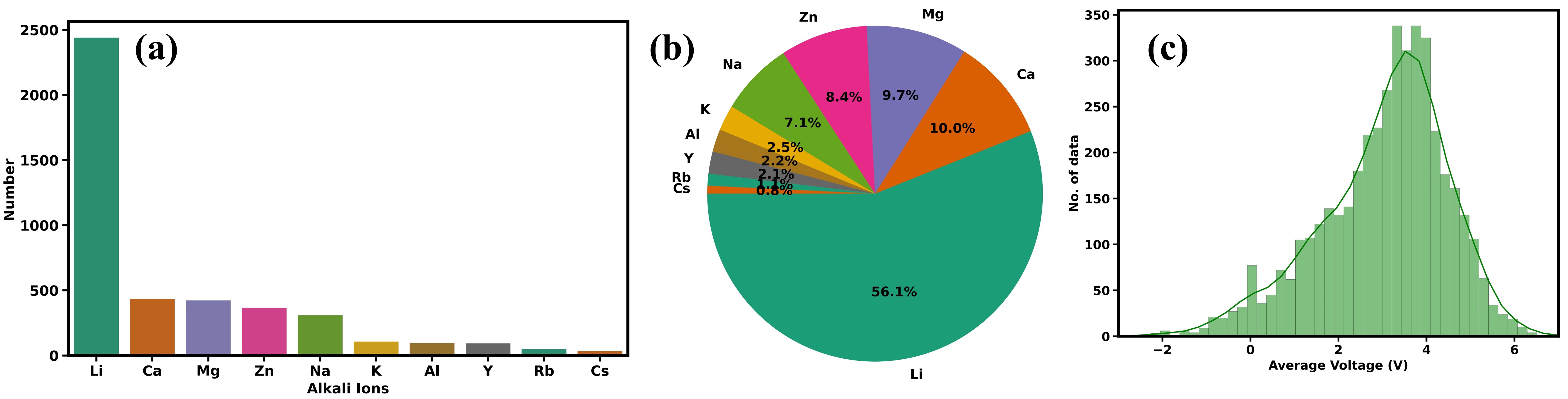}
    \caption{Statistical distribution of (a) 10 different alkali working metal-ion battery (b) \% population of metal-ion in the battery data set and (c) Average voltage distribution.}
    \label{fig:fig_1}
\end{figure}

In this work, we have collected all the data on metal-ion batteries from the Open Access Materials Project database (MP, v2022.10.28) using Pymatgen Material Genomes tools (pymatgen)\cite{jain2013commentary,ong2015materials}. We extracted 4351 data for all the metal-ion batteries with DFT computed voltage and their corresponding battery features with structure. The distribution of the number of 10 different alkali metal-ion batteries is shown in Figure \ref{fig:fig_1} (a) on a bar graph, and their corresponding weight percentage (pie graph) in the metal-ion battery domain is given in Figure \ref{fig:fig_1}(b). It clearly shows that the lithium metal-ion battery dominates the dataset when compared to other metal-ion batteries. In addition, we plot the average voltage distribution of the data set (shown in Figure \ref{fig:fig_1} (c)), with values ranging from 0 to 6.5 V, and this indicates the number of available data distributions with certain voltage.

\subsection{DNN architecture and training}
\begin{figure}[htp]
    \centering
    \includegraphics[width=15cm]{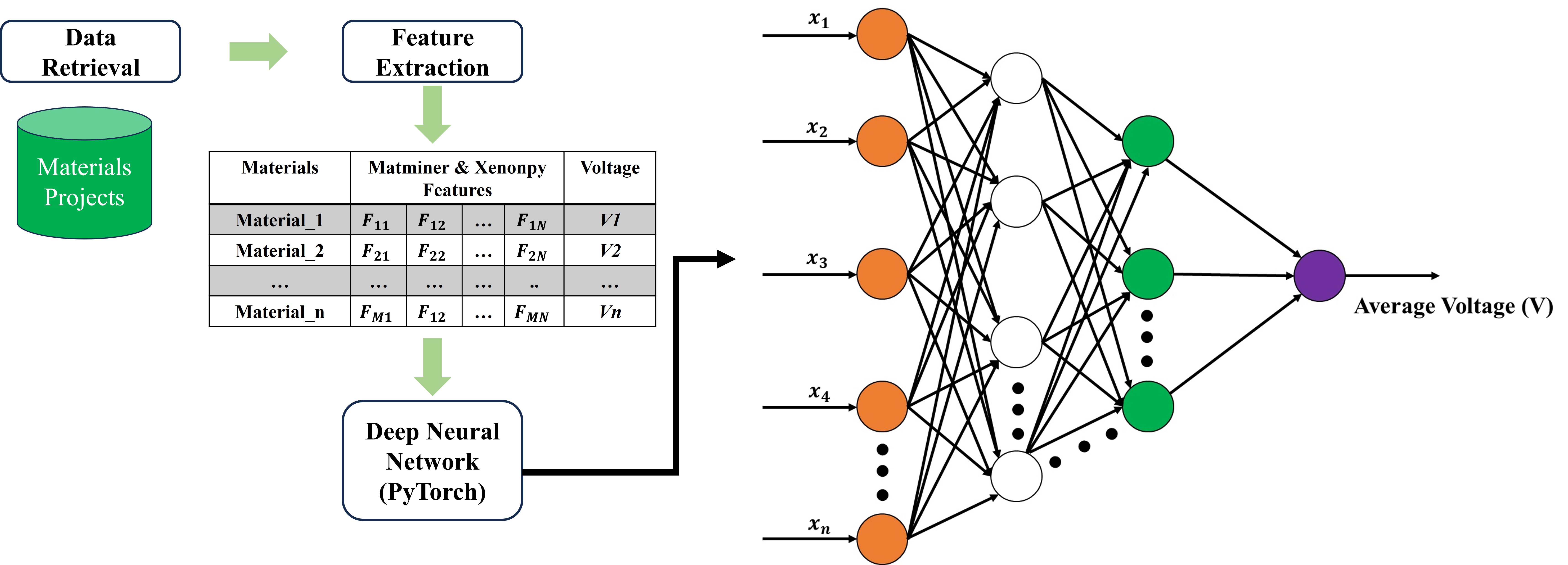}
    \caption{Workflow and DNN model architecture.}
    \label{fig:fig_2}
\end{figure}

The features of each charge and discharge electrode of all battery materials in the MP dataset were constructed using different composition-based feature generation tools such as Matminer\cite{ward2018matminer} and Xenonpy \cite{liu2021machine,pham2017machine}. In addition to key battery-related features, we also incorporate the statistical elemental properties of all constituent elements in each electrode composition. This approach helps capture the unique fingerprint of each electrode material within the dataset, enhancing the model's ability to differentiate and predict material properties accurately. The statistical elemental characteristic will be different for a specific electrode material, and by considering all statistical variables (such as avg, min, max, and var) we have generated almost 262 features (see Figure \ref{fig:fig_2}) and used them as input for the DNN model. Out of 262 features, we have generated 232 features from Xenonpy, and others are extracted from Matminer and battery explorer of MP. The components of a feature vector are diverse in terms of magnitude, which can range from small fractions to a few thousands. Therefore, all input features were normalized for better and more efficient training of the DNN model, thereby avoiding bias of a particular feature over others based solely on their magnitude. The normalization was performed by effectively scaling all the inputs to be between -1 and 1, and it was carried out while training our model and it is not done for the target value. 
The list of statistical features used to develop the DNN model is provided in Table S1 of the Supporting Information (SI).

%\subsection{Model Architecture}
To predict the average voltage of any electrode material (cathode or anode), we implemented a DNN based model on the battery electrode material properties. The DNN architecture was designed and implemented using PyTorch\cite{paszke2017automatic,paszke2019pytorch}, a flexible deep learning framework. The model consisted of the following layers: input layer, hidden layers, dropout layers, and output layer. The input layer accepts a feature vector derived from material properties, where the feature size corresponds to the dimensions of the pre-processed dataset. The network included 5 fully connected hidden layers, each with 262 neurons. Nonlinear activation functions, specifically LeakyReLU (Leaky Rectified Linear Unit), were applied to introduce nonlinearity and also prevent overfitting. Regularization of dropouts was incorporated with a dropout probability of 0.2 to prevent overfitting and improve generalization by randomly dropping neurons during training. The output layer consisted of a single neuron with a linear activation function, providing the predicted value of the average voltage. The performance of the ML models has been tested using different error metrics, mean absolute error (MAE), and mean square error (MSE). The MAE and MSE are defined by the following equations:

\begin{equation}
    MAE = \frac{1}{N}\sum_{i=1}^{N}(|V_{i}^{DFT} - V_{i}^{ML}|) \label{eqn: eqn_1}
\end{equation}

\begin{equation}
    MSE = \frac{1}{N}\sum_{i=1}^{N}(V_{i}^{DFT} - V_{i}^{ML})^{2} \label{eqn: eqn_2}
\end{equation}

%\begin{equation}
%    RMSE = \sqrt{\frac{1}{N}\sum_{i=1}^{N}(V_{i}^{DFT} - V_{i}^{ML})^{2}} \label{eqn: eqn_3}
%\end{equation}

where $V_{i}^{DFT}$ represents the voltage calculated from DFT and $V_{i}^{ML}$ represents the predicted voltage from machine learning, for the given $i^{th}$  battery sample and $N$ total number of samples in the dataset.
The MSE loss was used to quantify the prediction errors. MSE was chosen for its sensitivity to large errors, ensuring that the model minimizes significant deviations in voltage predictions. The Adam optimizer\cite{kingma2014adam} was utilized for its efficiency and adaptive learning rate capabilities. The learning rate and weight decay were set to 0.0001 and 0.001. 

For training, the data set was split into training and testing subsets, with 80\%  of the data allocated for training and the remainder for testing. The model was trained over 1000 epochs with a batch size of 64, balancing computational efficiency and convergence. A StepLR Learning Rate Scheduler was employed to dynamically adjust the learning rate, ensuring better convergence. During training, both training and validation losses were monitored to ensure that the model's performance improved without overfitting. A residual plot showed that the prediction errors were well distributed around zero, with no significant bias observed. The network architecture was defined using PyTorch's torch.nn.Module. The model parameters were updated iteratively using backpropagation, and gradients were computed using PyTorch's autograd functionality. After each epoch, the model was evaluated on the validation set to monitor the generalization performance. The DNN demonstrated R$^{2}$ value of 0.995 and 0.906 for training and testing with a MSE value of 0.009 and 0.182 indicating its effectiveness in predicting the mean voltage of battery materials. The model predictions closely matched the actual values (Figure \ref{fig:fig_4}), highlighting its ability to learn complex relationships between the input features and the target variable.

\subsection{First-principles DFT \& Boltzmann transport calculations}

All DFT calculations were performed using the Vienna Ab initio Simulation Package (VASP)\cite{kresse1999ultrasoft,kresse1996efficient} with spin-polarized ferromagnetic ordering. Formation energy calculations used the generalized gradient approximation (GGA) with the Perdew–Burke–Ernzerhof (PBE) exchange–correlation functional\cite{perdew1996generalized}. For voltage calculations, we applied the Hubbard correction parameters (U) to account for self-interaction errors, and the DFT-D3 method\cite{grimme2010consistent} was used to incorporate dispersion effects for all compounds of interest as the van der Waals correction is important for describing layer interactions.
The U parameters used for Fe, Mn, Ni, Co, Cr and Cu were 3.9, 3.9, 6.2, 3.32, 3.7, and 5.5 eV, respectively, as obtained from Ref \cite{wang2006oxidation,wei2020first}. A kinetic energy cutoff of 500 eV with k-point mesh of 11 × 7 × 3 for O3 and 3 x 3 x 2 for P2-type materials was found to be sufficient to achieve convergence. During structural optimization, the atomic positions were allowed to relax until the Hellmann–Feynman forces were reduced below 0.01 eV/\AA. Electronic minimization was performed with a convergence criterion of $10^{-6}$ eV. The new composition of the charge balance structure for the O3 and P2-type was built using the Supercell package \cite{okhotnikov2016supercell}. To generate the new compositions, here we have considered a 2 x 3 x 1 supercell lattice for O3-type materials and a 3 x 3 x 1 supercell lattice for P2-type materials.

The formation energy per formula unit (eV/f.u.) has been calculated using the formula given below \cite{emery2017high,mishra2025designing}: 

\begin{equation}
    E_f = E(Na_xM_1^aM_2^bM_3^cO_2) - x\mu_{Na} - a\mu_{M_1} - b\mu_{M_2} - c\mu_{M_3} - \mu_{O_2} \label{equ: equ_3}
\end{equation}
where $a + b + c =1$ and $M_1^a$, $M_2^b$, and $M_3^c$ represents the three different stoichiometry of metallic cations and their corresponding chemical potentials $\mu_{M_1}$, $\mu_{M_2}$, and $\mu_{M_3} $ respectively. Also, $\mu_{Na}$ and $\mu_{O_2}$ represents the chemical potential of Na bulk and O$_2$ molecule.

The voltage (V) of the newly designed compositions was calculated using the general formula\cite{gwon2009combined} for multicomponent layered metal oxide which is 

\begin{equation}
    V = -\frac{E_{Na_{x_2}MO_2} - E_{Na_{x_1}MO_2} - (x_2 - x_1)E_{Na} }{(x_2 - x_1)} \label{eqn: equ_4}
\end{equation}

where M is the multications of different stoichiometry of $M_1^a$  ,$M_2^b$ and  $M_3^c$; $E_{Na}$  is total energy of the bulk Na and $x_1$ and $x_2$ are the different concentrations of sodium (Na) during sodiation and desodiation.

To obtain electrical conductivity, we performed semi-classical Boltzmann transport calculations within the constant relaxation time approximation as implemented in the BoltzTraP code \cite{madsen2006boltztrap}. In this calculation, a denser k-mesh is adopted to obtain the density of states, for instance, 19$\times$15$\times$11 k-mesh is used for O3-type  and 18$\times$18$\times$16 for P2-type compositions. For P2-type composition with x=0.67, 16$\times$16$\times$12 k-mesh is used as it generates more number
of irreducible k-points (1540) than 18x18x16 k-mesh used for x=1 and x=0.33 (1379).

\section{Results and discussion}
\subsection{DNN performance and cross-validation}
\begin{figure}[htp]
    \centering
    \includegraphics[width=15cm]{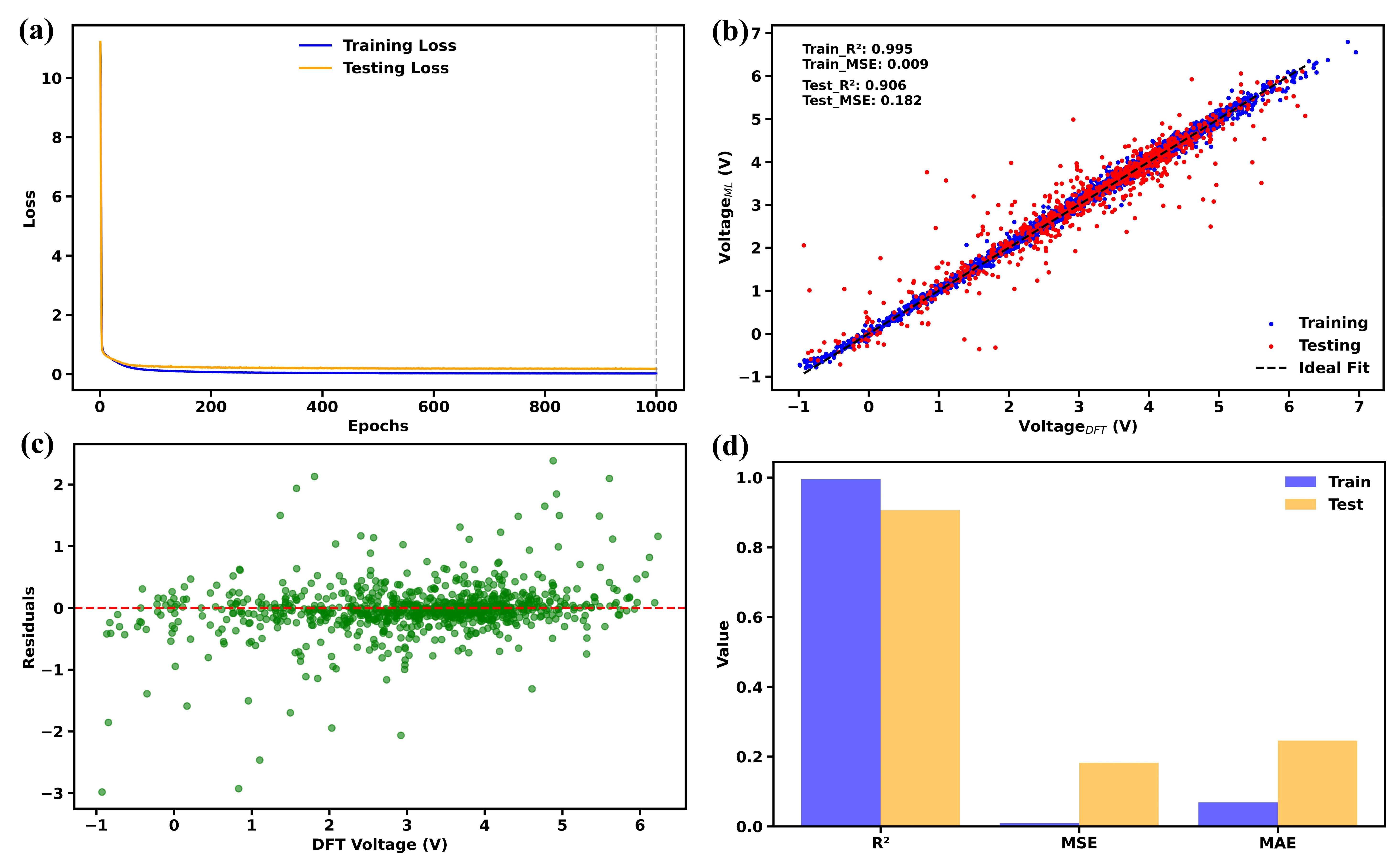}
    \caption{Training and testing of DNN model (a)  loss with epochs, (b) DFT voltage and ML predicted voltage, (c) residuals of actual and predicted voltage, and (d) metric value.}
    \label{fig:fig_3}
\end{figure}

The MSE loss during the model training was evaluated as shown in Figure \ref{fig:fig_3}(a). At the beginning of training (epochs close to 0), both training and testing losses are high, which is expected because the model starts with randomly initialized weights. It has not yet learned to map the input features to the target output effectively. In the first few epochs, both the training and the testing losses drop steeply. This indicates that the model is learning rapidly and effectively by adjusting its weights to minimize error. As training progresses, the loss values for both training and testing data begin to plateau. This suggests that the model has learned the underlying patterns in the data and is approaching convergence. Training loss and testing loss closely follow each other throughout the training process. This is a strong indicator that the model is generalizing well to unseen data and is not overfitting. At the final epochs (e.g., around 1000), both training and testing losses remain low and stable, further confirming that the model has achieved a good fit without over-fitting or under-fitting. The convergence of training and training losses demonstrates the model's capability to learn and generalize effectively. The lack of significant divergence between the two curves suggests that the chosen hyperparameters (learning rate, architecture, etc.) are appropriate. The alignment of the test loss with the training loss highlights that the model is not overly complex and has a good balance of bias and variance.  Here, the R$^{2}$ score and error metrics (MSE or MAE) were initially evaluated using a single train-test split. However, to ensure superior generalization and model reliability, we have also performed a 10-fold cross-validation, as discussed below.
% Insert figure here for loss 

The R$^{2}$ and MSE values for training and testing presented in Figure \ref{fig:fig_3} (b), show a near perfect R$^{2}$ and a very low MSE for the training set, which indicate that the model has successfully learned the patterns in the training data. The test R$^{2}$ value of 0.906 shows that the model captures most of the variance in the test data.  Some scatter points deviate significantly from the ideal-fit line, especially at higher actual voltage values. These deviations indicate instances where the model struggles to predict accurately. The higher MSE in the test set (0.182 vs 0.009, Figure \ref{fig:fig_3} (d)) reflects that very small prediction errors could be present in unseen data, which is common in machine learning models. The model demonstrates strong learning in the training set and reasonable generalization in the test set, as evidenced by the high test R$^{2}$ and close alignment of many points with the ideal fit line ($y=x$). Minor overfitting may occur, but the model can still make accurate predictions for most test data. From the MSE and MAE values (0.182 and 0.24V, respectively), it can be concluded that our DNN model shows excellent performance compared to previously reported ML models, with the respective MAE \cite{moses2021machine,joshi2019machine,louis2022accurate} shown in Table \ref{tbl:table1}. The error bar graph in Figure \ref{fig:fig_3}(d) compares the mean error metrics of our model. It shows a smaller MAE with respect to the MAE values previously reported by \citeauthor{joshi2019machine}\cite{joshi2019machine} and \citeauthor{moses2021machine}\cite{moses2021machine}. The residuals appear fairly symmetric around the zero line (Figure \ref{fig:fig_3} (c)), suggesting that the model does not have a significant bias towards overestimating or underestimating the actual values. Most residuals are clustered close to the zero line, indicating that the model's predictions are generally accurate. %However, some residuals deviate significantly, especially for higher actual voltage values (e.g., above 4V), suggesting that the model struggles with accurate predictions for high-voltage cases. 

\begin{table}
  \caption{Comparison of our DNN model and other models}
  \label{tbl:table1}
  \begin{tabular}{llll}
    \hline
    Models  & R$^2$ & MAE  & Source \\ %& V$_{DNN}$ & $\Delta V1$ & $\Delta V2$ \\
    \hline
    DNNs   & 0.81 & 0.43 & \citeauthor{joshi2019machine}\cite{joshi2019machine} \\
    DNNs   & 0.83 & 0.39 & \citeauthor{moses2021machine}\cite{moses2021machine} \\
    GNNs   & - - - & 0.34 & \citeauthor{louis2022accurate}\cite{louis2022accurate} \\
    DNNs   & 0.91 & 0.24 & \textbf{Present work} \\
    \hline
  \end{tabular}
\end{table}

\begin{figure}[htp]
    \centering
    \includegraphics[width=15cm]{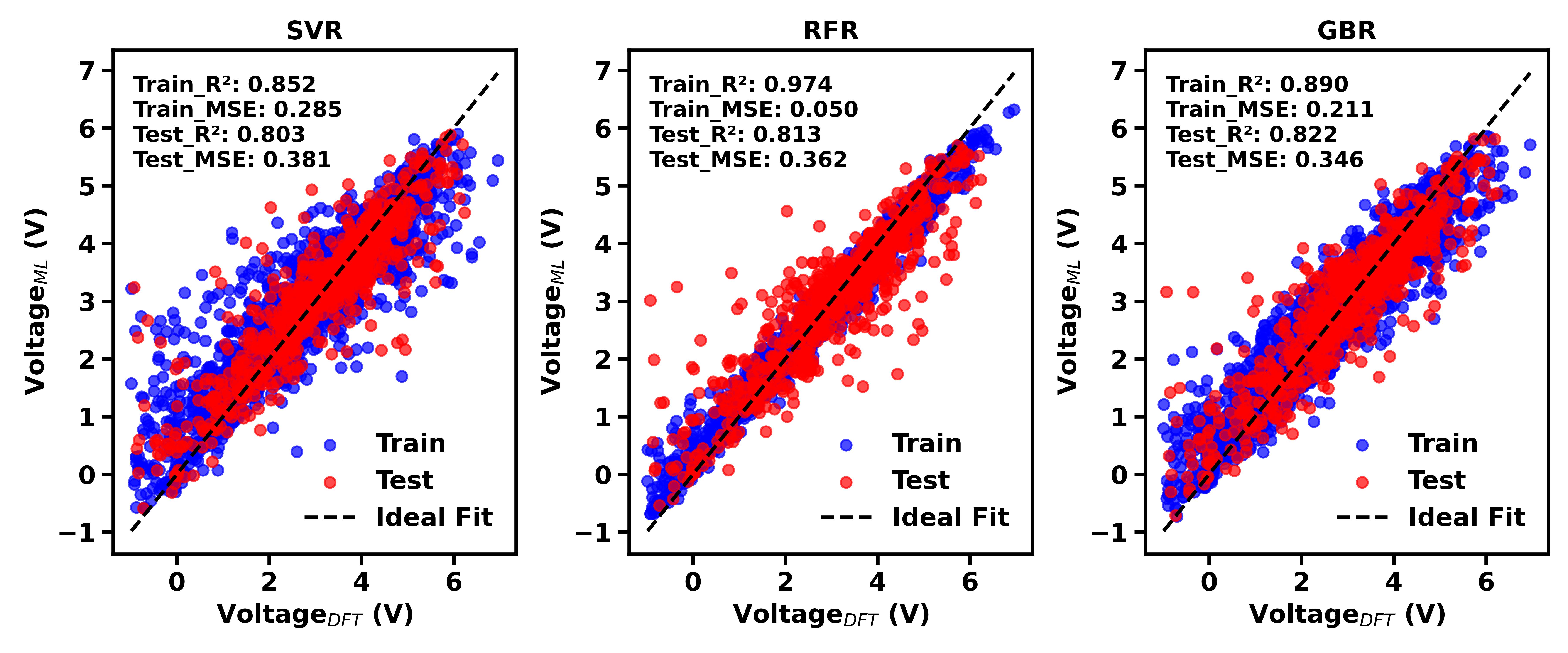}
    \caption{Performance comparison of traditional machine learning models SVR, RFR, and GBR for predicting the average voltage of battery materials.}
    \label{fig:fig_33}
\end{figure}

In addition to the DNN model, we trained several traditional regression-based machine learning models, including random forest regression (RFR), support vector regression (SVR), and gradient booster regression (GBR), to predict the average voltage of battery materials. Figure \ref{fig:fig_33} illustrates the performance of these models in the training and testing phases, with the corresponding coefficient of determination R$^{2}$ and the MSE values. Among the models evaluated, RFR achieved the highest predictive accuracy on the test dataset, yielding a R$^{2}$ value of 0.813.
Despite the relatively strong performance of the RFR model compared to SVR and GBR, it was outperformed by the DNN model, as shown in \ref{fig:fig_3}(b). The DNN model, implemented using the PyTorch framework, demonstrated a superior generalization capability, capturing complex nonlinear relationships in the dataset more effectively. In particular, while traditional machine learning models rely on manually engineered features and predefined structures, deep learning approaches leverage hierarchical feature extraction, allowing for improved representation learning. The significantly higher predictive accuracy of the DNN model underscores its effectiveness in voltage prediction tasks for battery materials, particularly when dealing with high-dimensional feature spaces.

\begin{figure}[htp]
    \centering
    \includegraphics[width=15cm]{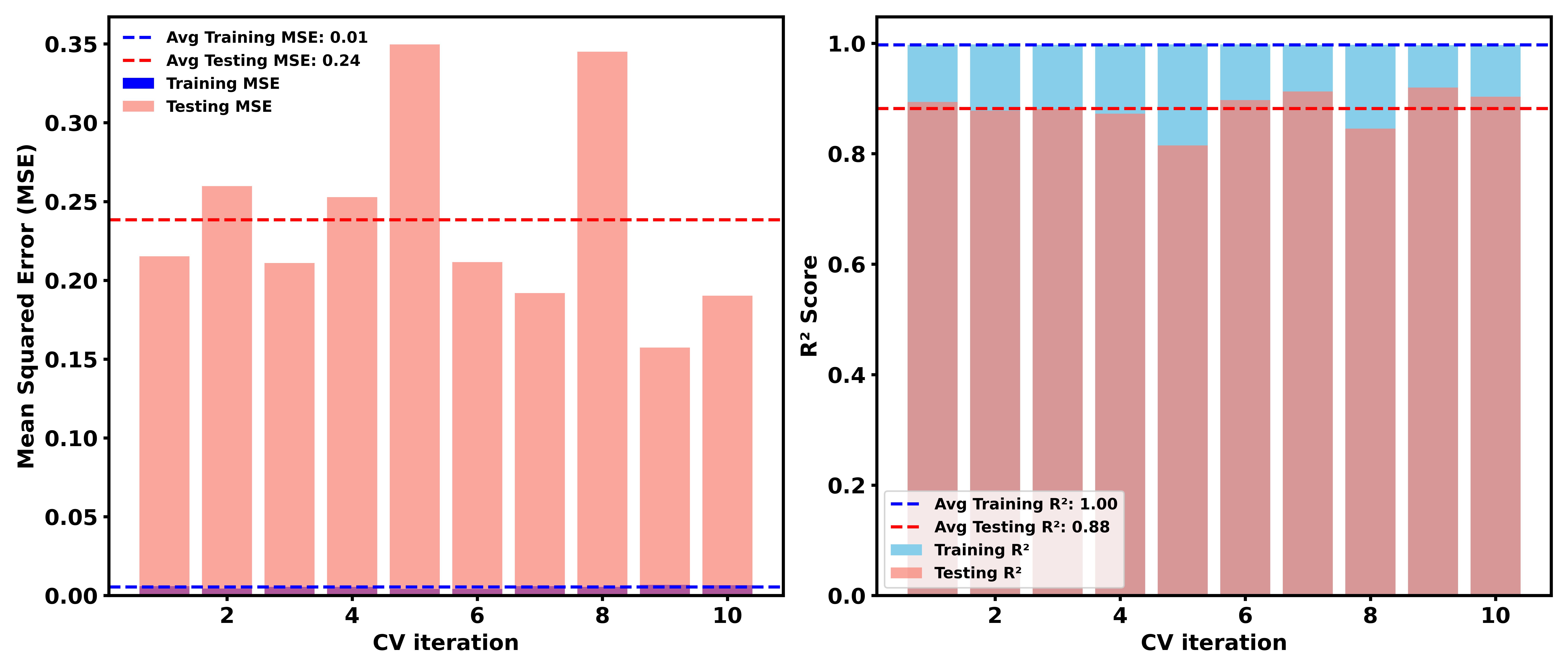}
    \caption{Cross-validation performance of the DNN model: (Left) MSE for training and testing sets across 10 folds, with average values indicated by dashed lines. (Right) R² scores for training and testing sets, demonstrating model generalization.}
    \label{fig:fig_cv}
\end{figure}

%\subsection{Cross-Validation of the DNN Model}
 Figure \ref{fig:fig_cv} present the results of a 10-fold cross-validation analysis, evaluating both the MSE and the R$^{2}$ score for the training and testing sets. The low training MSE and the high training R$^{2}$ (close to 1.0) indicate that the model effectively captures the underlying data patterns, while the consistently high testing R$^{2}$ ($\sim$ 0.88) and stable testing MSE confirm its strong generalizability. To avoid overfitting, we employed feature selection techniques, regularization, and hyperparameter tuning to optimize the complexity of the model. In addition, the model was evaluated on an independent test set, ensuring its robustness against unseen data. These results demonstrate that the model achieves a balance between high predictive accuracy and generalization performance, making it well suited for battery voltage predictions.

%\subsection{Validation of the DNN model}
%Different ML algorithms performance was assessing to achieve the goal of voltage prediction. Herein we have employed four  regression based machine learning algorithms such as RFR, GBR, SVR, and Decision tree to predict the average voltage of electrode materials. We have used mean absolute error (MAE), mean squared error (MSE) and Root mean square error (RMSE) as a metric to measure the performance of our ML model as implemented in the Scikit-Learn machine learning library(). After performing 10 cross-validation we have shown the mean and standard deviation (SD) of all the MAE, MSE, and RMSE metric value in the \textbf{Table~\ref{tbl:metric error}}. From the MAE value of 0.09 $\pm$ 0.01 V suggests that the RFR ML model showing the excellent performance as compared to the others ML models with best optimize parameters. The error bar plot in \textbf{Figure \ref{fig:metric errors}} compare the mean error metrics of all the ML models and our RFR model showing the smaller MAE value compared to the previously reported MAE value by \citeauthor{joshi2019machine}\cite{joshi2019machine} and \citeauthor{moses2021machine}\cite{moses2021machine}

\begin{figure}[htp]
    \centering
    \includegraphics[width=15cm]{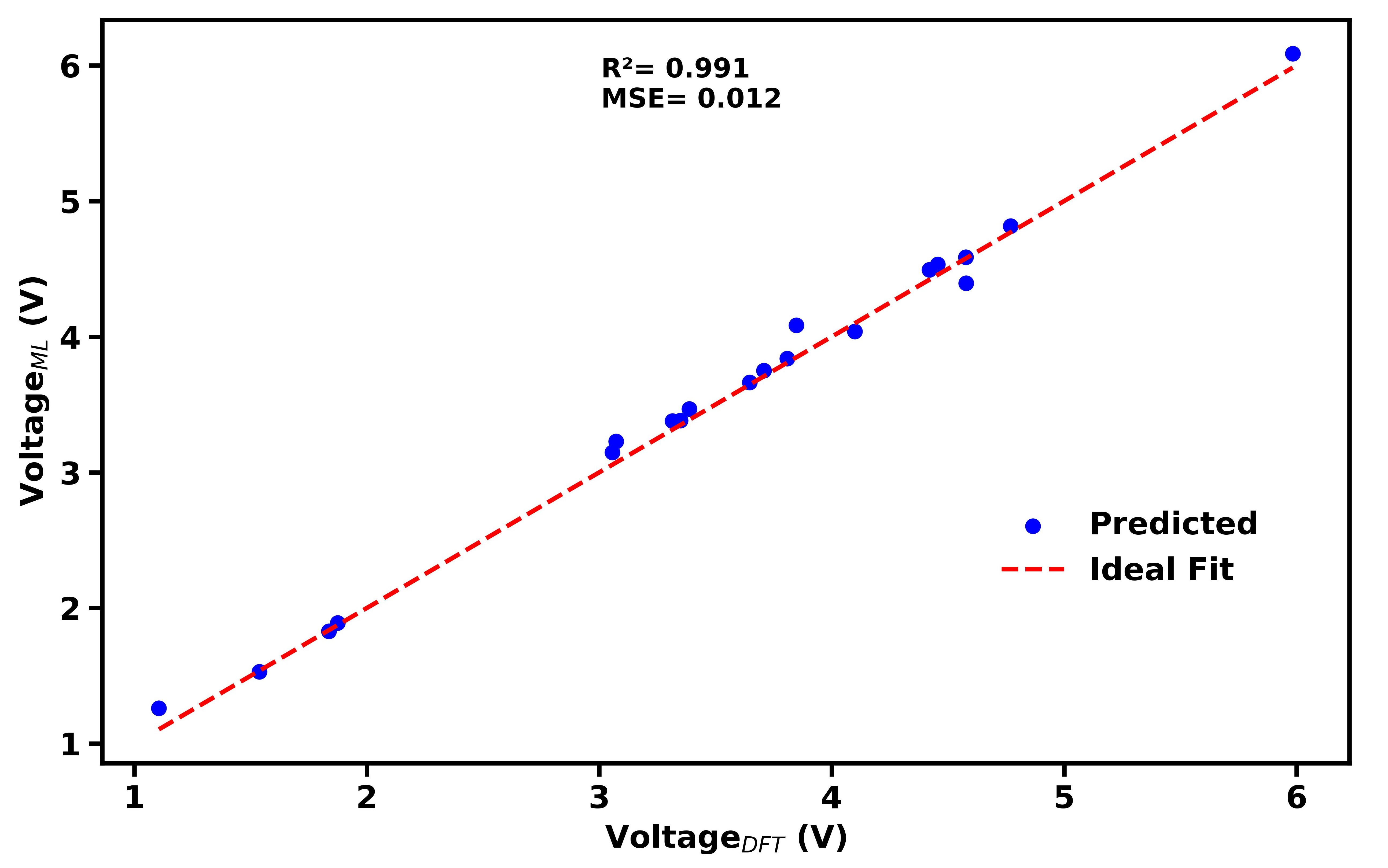}
    \caption{Model validation with DFT data of experimentally available electrode materials.}
    \label{fig:fig_4}
\end{figure}

Model validation is crucial for testing its excellent performance verification and evaluation. We validate our DNN model by considering experimentally observed and commercially available battery electrode material for both Li-ion and Na-ion batteries (see Table \ref{tbl:table2}). In Figure \ref{fig:fig_4}, we show the predicted voltages against the DFT-calculated voltages for various electrode materials, demonstrating the accuracy and reliability of the model. The plot shows a strong correlation between the predicted and DFT-calculated voltages, with an R$^2$ value of 0.991 and an MSE of 0.012V. The agreement between the predicted and experimental voltages is also excellent, with an average absolute deviation of 0.05V. This demonstrates the ability of the model to accurately predict voltages for various electrode materials, which is essential for the design of high-performance batteries.

The list of materials used for the validation is shown in Table \ref{tbl:table2} along with voltage data obtained from DFT, our DNN model, and the experiment. The absolute deviation of the predicted voltage of our model with the experimental and DFT voltages is evaluated through the following expressions: $\Delta V_1 = |V_{DFT} - V_{DNN}|$ and $\Delta V_2 = |V_{expt.} - V_{DNN}|$, respectively, and is tabulated in Table \ref{tbl:table2}.

\begin{table}
  \caption{DNN predicted vs experimental and DFT voltages and their deviation}
  \label{tbl:table2}
  \begin{tabular}{lllllll}
    \hline
    Materials  & V$_{expt.}$ & V$_{DFT}$  & V$_{DNN}$ & $\Delta V_1$ & $\Delta V_2$ \\
    \hline
    LiCoO$_{2}$   & 4.1\cite{amatucci1996coo2} & 3.808 & 3.803 & 0.005 & 0.297  \\
    LiFePO$_{4}$  & 3.5\cite{delmas1999lithium} & 3.847 & 3.845 & 0.002 & 0.345 \\
    LiMnPO$_{4}$  & 4.1\cite{li2002limnpo4} & 4.578 & 4.207 & 0.371 & 0.107 \\
    LiCoPO$_{4}$  & 4.8\cite{amine2000olivine} & 4.770 & 4.753 & 0.017 & 0.047 \\
    LiNiO$_{2}$   & 3.85\cite{delmas1999lithium} & 4.099 & 4.079 & 0.020 & 0.229 \\
    LiMn$_{2}$O$_{4}$ & 4.15\cite{priyono2019electrochemical} & 4.577 & 4.545 & 0.032 & 0.395 \\
    LiTiS$_{2}$   & 2.1\cite{whittingham1976electrical} & 1.875 & 1.866 & 0.009 & 0.234 \\
    LiNiPO$_{4}$  & 5.3\cite{wolfenstine2005ni3+} & 5.983 & 5.957 & 0.026 & 0.657 \\
    Li$_{3}$V$_{2}$(PO$_{4}$)$_{3}$ & 3.8\cite{gaubicher2000rhombohedral} & 3.708 & 3.716 & 0.008 & 0.084 \\
    Na$_{3}$V$_{2}$(PO$_{4}$)$_{3}$ & 3.4\cite{gaubicher2000rhombohedral} & 3.349 & 3.361 & 0.012 & 0.039 \\
    NaCoO$_{2}$   & 2.8\cite{delmas1981electrochemical,berthelot2011electrochemical} & 3.056 & 3.043 & 0.013 & 0.243 \\
    NaNiO$_{2}$   & 3.0\cite{braconnier1982etude} & 3.315 & 3.277 & 0.038 & 0.277 \\
    NaTiO$_{2}$   & 1.5\cite{maazaz1983study} & 1.106 & 1.104 & 0.002 & 0.396 \\
    NaFePO$_{4}$  & 3.0\cite{moreau2010structure} & 3.072 & 3.046 & 0.026 & 0.046 \\
    \hline
  \end{tabular}
\end{table}

 %Our model is developed based on DFT data and all there features therefor it is clear that the deviation ($\Delta V_1$) between our DNN model predicted voltage and DFT voltage is very small almost every validating material. Whereas large deviation ($\Delta V_2$) observed with the experimental value, and it is absolutely true as the voltage of any battery materials also depends on some experimental conditions (like humidity, temperature, pressure, etc.) which is not present in our DFT database. However our model still can predict the experimental value with minimal error as we have shown few cases presented in the Table \ref{tbl:table2}. 

 Our model is developed based on DFT data from battery materials and its associated features, ensuring that the deviation ($\Delta V_1$) between the predicted voltage of our DNN model and the calculated voltage by DFT remains minimal for nearly all validation materials. In contrast, a larger deviation ($\Delta V_2$) is observed when comparing the predicted values with the experimental data. This discrepancy is expected because the voltage of battery materials is influenced by various experimental conditions, such as humidity, temperature, and pressure, factors that are not considered in the DFT data set. Despite these challenges, our model demonstrates the ability to predict experimental voltages with minimal error, as evidenced by several cases presented in Table \ref{tbl:table2}.

\subsection{Feature importance}
\begin{figure}[htp]
    \centering
    \includegraphics[width=15cm]{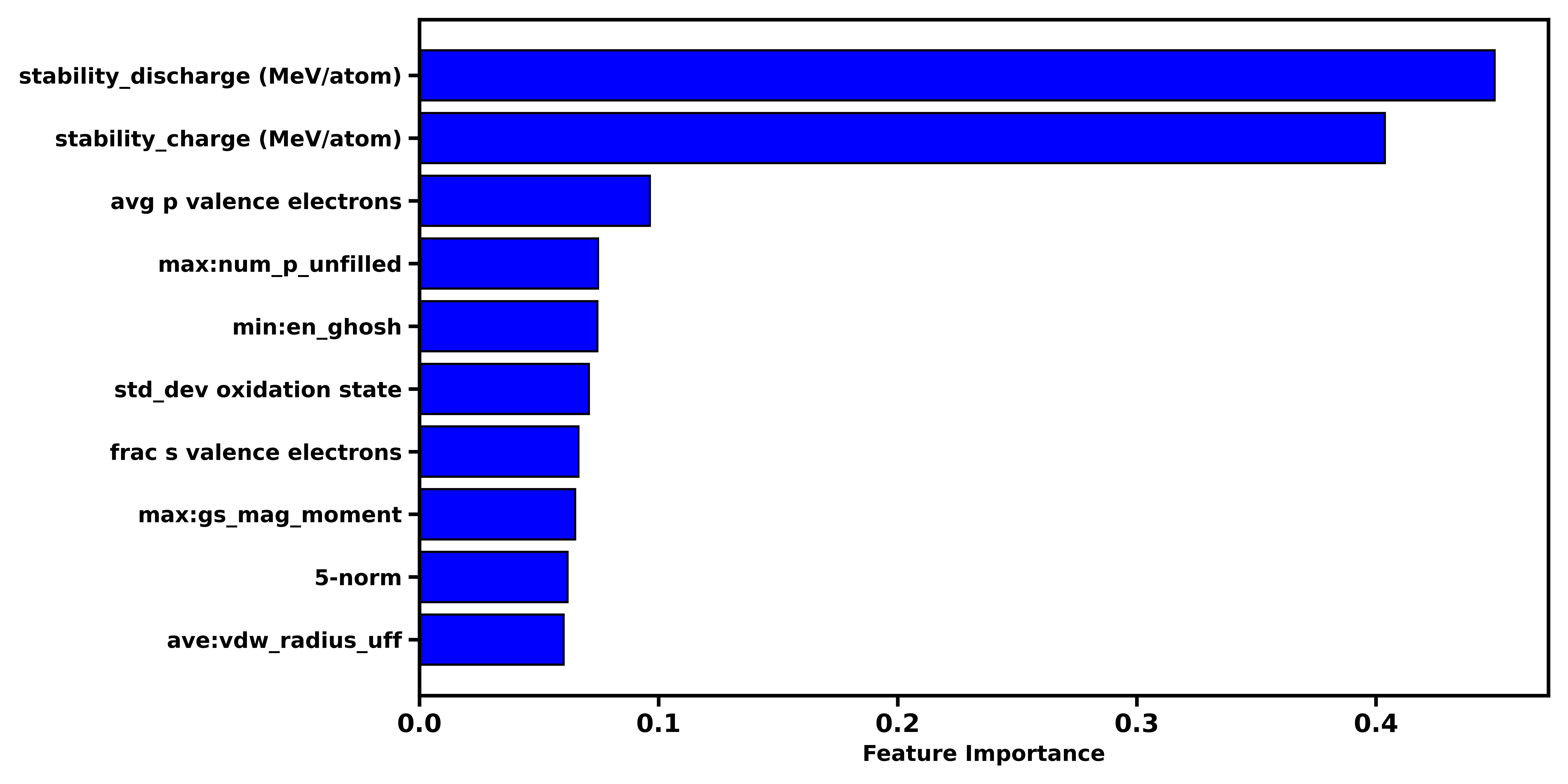}
    \caption{Key features affecting electrochemical performance.}
    \label{fig:fig_imp_fea}
\end{figure}

Figure \ref{fig:fig_imp_fea} represents the analysis of the importance of characteristics for voltage prediction in battery materials that reveals that the most influential descriptors are related to the thermodynamic stability, electronic structure and redox behavior of compounds. The top two features, stability\_discharge (MeV/atom) and stability\_charge (MeV/atom), indicate the energetic stability of the material in both charge and discharge states. These features are crucial because they reflect the material’s ability to reversibly intercalate and deintercalate ions without significant structural degradation, directly affecting the voltage profile \cite{liu2016understanding}. The average p-valence electrons and the maximum number of unfilled p orbitals (max:num\_p\_unfilled) provide insight into the electronic structure, particularly the capacity of the material to accommodate electron transfer during redox reactions\cite{saubanere2014intuitive}. The minimum energy of the Ghosh electronegativity scale \cite{ghosh2005new,ghosh2009electronegativity} (min: en\_ghosh) captures the material's ability to stabilize anion-cation interactions, which is critical for determining the redox potential\cite{sharma2023computational}. Furthermore, the standard deviation of the oxidation states reflects the variability in the oxidation behavior between the constituent elements, influencing the charge transfer dynamics\cite{louis2022accurate}. Features such as fraction of s valence electrons, maximum ground-state magnetic moment (max:gs\_mag\_moment), and 5-norm\cite{moses2021machine} (a measure of geometric or electronic structure variability) further contribute by characterizing bonding nature, magnetic properties, and structural stability, respectively. Lastly, the average van der Waals radius (ave:vdw\_radius\_uff) highlights the role of atomic size in determining ion mobility within the material\cite{sotoudeh2024ion}. These features collectively define the electrochemical and voltage characteristics of battery materials, which aids in the rational design of high-performance electrodes.

\subsection{ML-guided cathode design}
\begin{figure}[htp]
    \centering
    \includegraphics[width=15cm]{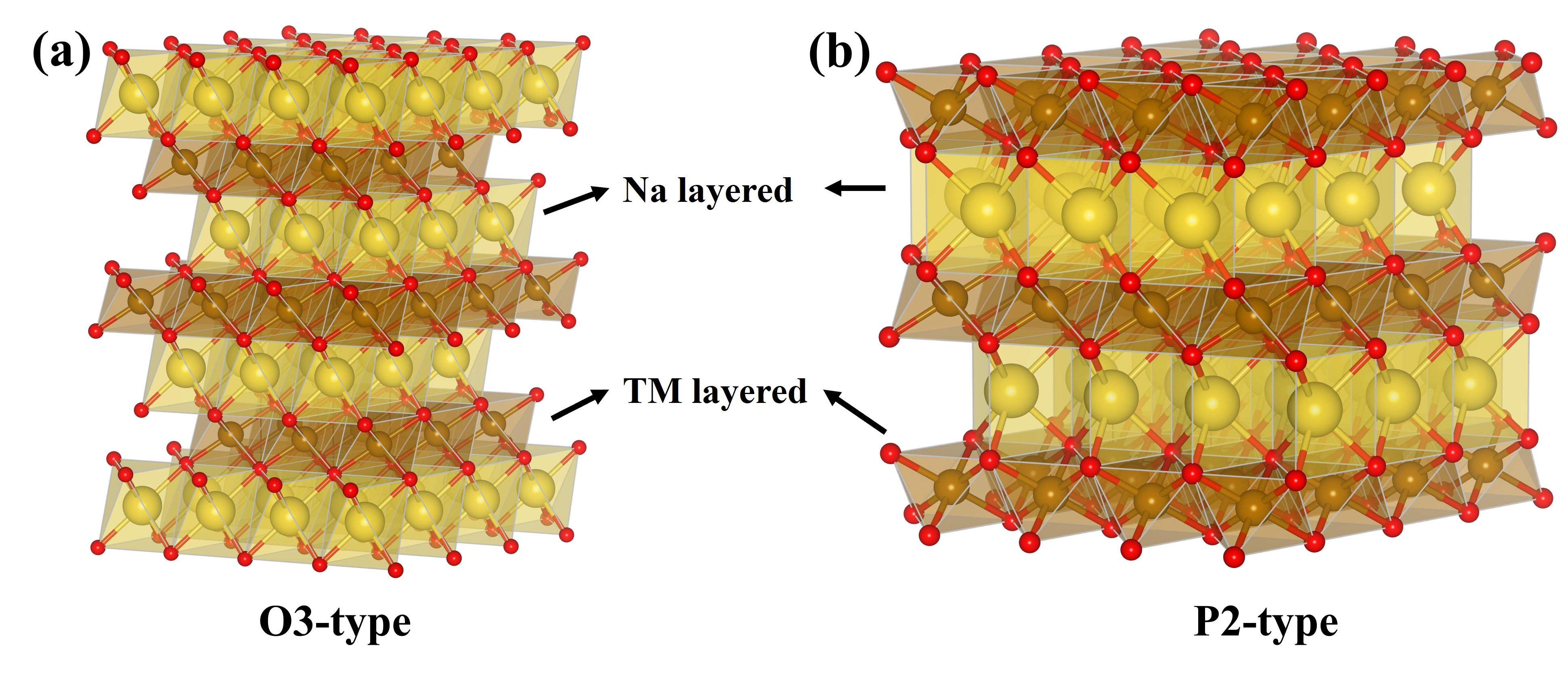}
    \caption{Structure of transition metal layered oxide of (a) O3 and (b) P2-type.}
    \label{fig:fig_6}
\end{figure}

The design of new Na-ion battery cathode materials was carried out using template crystal structures of transition-metal layered oxides with the general formula NaMO$_2$, where M represents transition or post-transition metals. These layered oxides serve as a fundamental structural framework as a result of their well-established electrochemical properties and stability in Na-ion battery applications.
In this study, we focus on two widely studied structural types of layered transition-metal oxides. O3 and P2-type materials (see Figure \ref{fig:fig_6}). The O3-type materials exhibit trigonal symmetry with the \( R-3m \) space group, while the P2-type materials possess hexagonal symmetry with the \( P6_3/mmc \) space group. These two structural configurations were carefully considered in the design of new compositions due to their distinct sodium diffusion pathways and electrochemical characteristics.

To generate novel compositions, we explored a wide range of 3d transition metals (Cr, Ti, V, Mn, Fe, Ni, Co, Cu, Zn) along with other selected metal species (Al, Mg, Sr), aiming to achieve an optimal balance of critical electrochemical properties. The primary selection criteria for these elements were phase stability, high voltage, long cycle life, efficient Na ion diffusion, and fast charging capability. The Supercell package \cite{okhotnikov2016supercell} was utilized to systematically explore and enumerate various combinations of elements within the NaMO$_2$ framework. This computational approach enabled the creation of an extensive library of potential compositions by generating numerous atomic substitutions within the host structure. The resulting data set includes a large number of unique chemical compositions for O3 and P2-type cathodes, each exhibiting distinct electrochemical and structural properties.

A comprehensive list of newly designed O3 and P2-type cathode materials is presented in Figure \ref{fig:fig_7}. The voltage predictions were obtained using our PyTorch-based DNN model, which has demonstrated superior accuracy in predicting the electrochemical performance of Na-ion cathodes. To further validate the stability and feasibility of these newly designed compositions, we performed first-principles DFT calculations. Tables \ref{tab:my_label} and \ref{tab:my_label_1} represent the newly designed material compositions with formation energy, assessing their thermodynamic stability and the likelihood of successful synthesis. A detailed discussion of the DFT results, including computed formation energies and structural analyses, is provided in the Supporting Information (SI) section.

\begin{table}
    \centering
    \begin{tabular}{cc}
    \hline
    O3-type Compositions  & Formation energy (eV/f.u.) \\ 
    \hline
    NaMn$_{0.61}$Ni$_{0.28}$Cr$_{0.11}$O$_2$  & -7.31 \\
    Na$_{0.67}$Mn$_{0.61}$Ni$_{0.28}$Cr$_{0.11}$O$_2$  & -6.69 \\
    Na$_{0.33}$Mn$_{0.61}$Ni$_{0.28}$Cr$_{0.11}$O$_2$  & -5.70 \\
    Na$_{0.83}$Mn$_{0.5}$Fe$_{0.28}$Cu$_{0.11}$Mg$_{0.11}$O$_2$  & -7.12 \\ Na$_{0.94}$Mn$_{0.5}$Fe$_{0.28}$Cu$_{0.11}$Mg$_{0.11}$O$_2$ &-7.33 \\
    Na$_{0.67}$Mn$_{0.5}$Ni$_{0.28}$Ti$_{0.11}$Cr$_{0.11}$O$_2$  & -7.03  \\
    NaMn$_{0.5}$Fe$_{0.33}$Cu$_{0.06}$Mg$_{0.11}$O$_2$ & -7.47 \\
    NaMn$_{0.5}$Fe$_{0.28}$Al$_{0.11}$Mg$_{0.11}$O$_2$ & -7.99 \\
    NaMn$_{0.5}$Fe$_{0.28}$Cu$_{0.11}$Al$_{0.11}$O$_2$ & -7.48 \\
    NaMn$_{0.5}$Fe$_{0.28}$Cu$_{0.11}$Cr$_{0.11}$O$_2$ & -7.11 \\
    \hline
    \end{tabular}
    \caption{New composition of O3-type materials for Na-ion battery}
    \label{tab:my_label}
\end{table}

\begin{table}
    \centering
    \begin{tabular}{cc}
    \hline
    P2-type Compositions  & Formation energy (eV/f.u.) \\ 
    \hline
    NaMn$_{0.5}$Fe$_{0.28}$Cu$_{0.22}$O$_2$  & -6.42 \\
    Na$_{0.89}$Mn$_{0.5}$Fe$_{0.28}$Cu$_{0.22}$O$_2$  & -6.58 \\
    Na$_{0.83}$Mn$_{0.5}$Fe$_{0.28}$Cu$_{0.22}$O$_2$  & -6.66 \\
    Na$_{0.72}$Mn$_{0.5}$Fe$_{0.28}$Cu$_{0.22}$O$_2$  & -6.82 \\ Na$_{0.67}$Mn$_{0.5}$Fe$_{0.28}$Cu$_{0.22}$O$_2$ & -6.89 \\
    Na$_{0.67}$Mn$_{0.5}$Fe$_{0.33}$Zn$_{0.17}$O$_2$  & -7.09  \\
    Na$_{0.67}$Mn$_{0.5}$Fe$_{0.33}$Mg$_{0.17}$O$_2$ & -7.06 \\
    Na$_{0.67}$Mn$_{0.5}$Fe$_{0.33}$Sr$_{0.17}$O$_2$ & -7.06 \\
    Na$_{0.67}$Mn$_{0.5}$Fe$_{0.33}$Ba$_{0.17}$O$_2$ & -7.02 \\
    Na$_{0.67}$Mn$_{0.5}$Fe$_{0.28}$Ni$_{0.22}$O$_2$ & -6.63 \\
    Na$_{0.67}$Mn$_{0.5}$Fe$_{0.28}$Co$_{0.17}$O$_2$ & -6.20 \\
    Na$_{0.67}$Mn$_{0.5}$Fe$_{0.28}$Ti$_{0.22}$O$_2$ & -5.99 \\
    Na$_{0.67}$Mn$_{0.5}$Fe$_{0.28}$Al$_{0.22}$O$_2$ &  -6.97 \\
    \hline
    \end{tabular}
    \caption{New composition of P2-type materials for Na-ion battery}
    \label{tab:my_label_1}
\end{table}

As summarized in Tables \ref{tab:my_label} and \ref{tab:my_label_1}, all investigated O3 polymorphs exhibit formation energies that are 0.3–1.0 eV f.u.$^{-1}$ lower (i.e., more negative) than those of their isochemical counterparts P2, indicating superior thermodynamic stability. This difference can be attributed to the distinct transition-metal–oxygen (TM–O) connectivity inherent to the two stacking sequences. In the O3 structure, each transition metal cation resides within a corner-sharing octahedron that also engages in extensive edge- and face-sharing with adjacent units. This results in a highly interconnected and six-fold coordinated TM-O framework. The enhanced orbital overlap within this dense network strengthens the overall TM–O bonding, making the creation of a sodium vacancy energetically more costly. In contrast, the P2 phases consist of prismatic TM–O units connected primarily through edge sharing, forming a more open and less interconnected framework. Consequently, the average TM–O interactions are weaker, leading to less negative formation energies.

%\begin{figure}[htp]
%    \centering
%    \includegraphics[width=14.5cm]{Figures/formation_energy_O3_P2_1.jpg}
%    \caption{Formation energy of newly design material composition of (a) O3 and (b) P2-type.}
%    \label{fig:fig_8}
%\end{figure} 

%Using DFT method (GGA+U) the predicted voltage by DNN model has been validated by considering one specific composition presented in green shown in Figure \ref{fig:fig_7}(a). The predicted voltage of the composition of  P2-Na$_{x}$Mn$_{0.5}$Fe$_{0.28}$Cu$_{0.22}$O$_{2}$ (where x is 1.00, 0.89, 0.83, 0.72, and 0.67) by the DNN model suggests that during charging or desodiation the voltage remain flat at around 4.2V upto x=0.67. The flat voltage profile (\ref{fig:fig_7}a) suggest that there is no phase transformation occurring during charging and discharging. Figure 8 represent the DFT calculated and DNN predicted voltage of P2-Na$_{x}$Mn$_{0.5}$Fe$_{0.28}$Cu$_{0.22}$O$_{2}$ and both are corroborating excellently well. All the structural detail of P2-Na$_{x}$Mn$_{0.5}$Fe$_{0.28}$Cu$_{0.22}$O$_{2}$ are discussed in the supporting information.
\subsection{Validation and analysis of ML-predicted materials}

\subsubsection{Voltage comparison (ML vs DFT)}
\begin{figure}[htp]
    \centering
    \includegraphics[width=15cm]{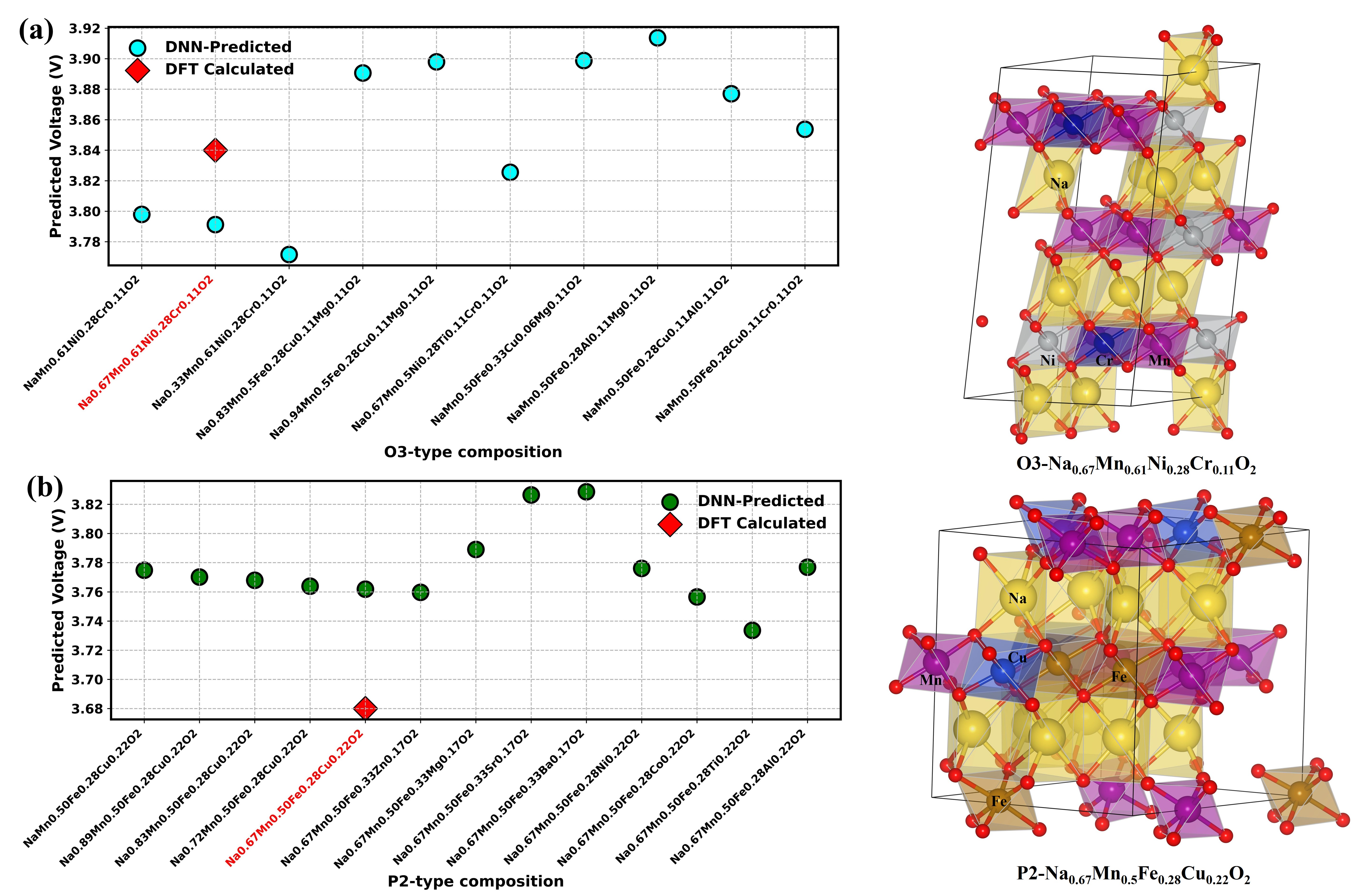}
    \caption{DNN predicted and DFT voltage of some specific cathode compositions for Na-ion batteries (a) O3-type and (b) P2-type.}
    \label{fig:fig_7}
\end{figure}

Using DFT calculations within the GGA+U framework, we have validated the voltage predictions generated by our DNN model for specific compositions: O3-Na$_{0.67}$Mn$_{0.61}$Ni$_{0.28}$Cr$_{0.11}$O$_{2}$ and P2-Na$_{0.67}$Mn$_{0.5}$Fe$_{0.28}$Cu$_{0.22}$O$_{2}$. These compositions, highlighted in red in Figure \ref{fig:fig_7}(a, b), were selected as representative validation points to avoid the computational expense of performing DFT calculations for all possible compositions. The DNN predicted voltages for O3-Na$_{0.67}$Mn$_{0.61}$Ni$_{0.28}$Cr$_{0.11}$O$_{2}$ and P2-Na$_{0.67}$Mn$_{0.5}$Fe$_{0.28}$Cu$_{0.22}$O$_{2}$ are 3.79 and 3.76V, respectively, while the corresponding DFT-calculated voltages are 3.84 and 3.68V (see SI), as shown in Figure \ref{fig:fig_7}(a, b). The minimal errors of only 1.3\% and 2.2\% demonstrate excellent agreement between the two methods. This strong correlation highlights the robustness and reliability of our machine learning model in accurately predicting voltages while significantly reducing the computational cost associated with traditional first-principles calculations. The consistency between the DFT and DNN results underscores the potential of our model as a powerful tool for screening and predicting the electrochemical properties of novel sodium-ion battery materials.

\subsubsection{DFT \& Boltzmann transport analysis}
\begin{figure}[htp]
    \centering
    \includegraphics[width=15cm]{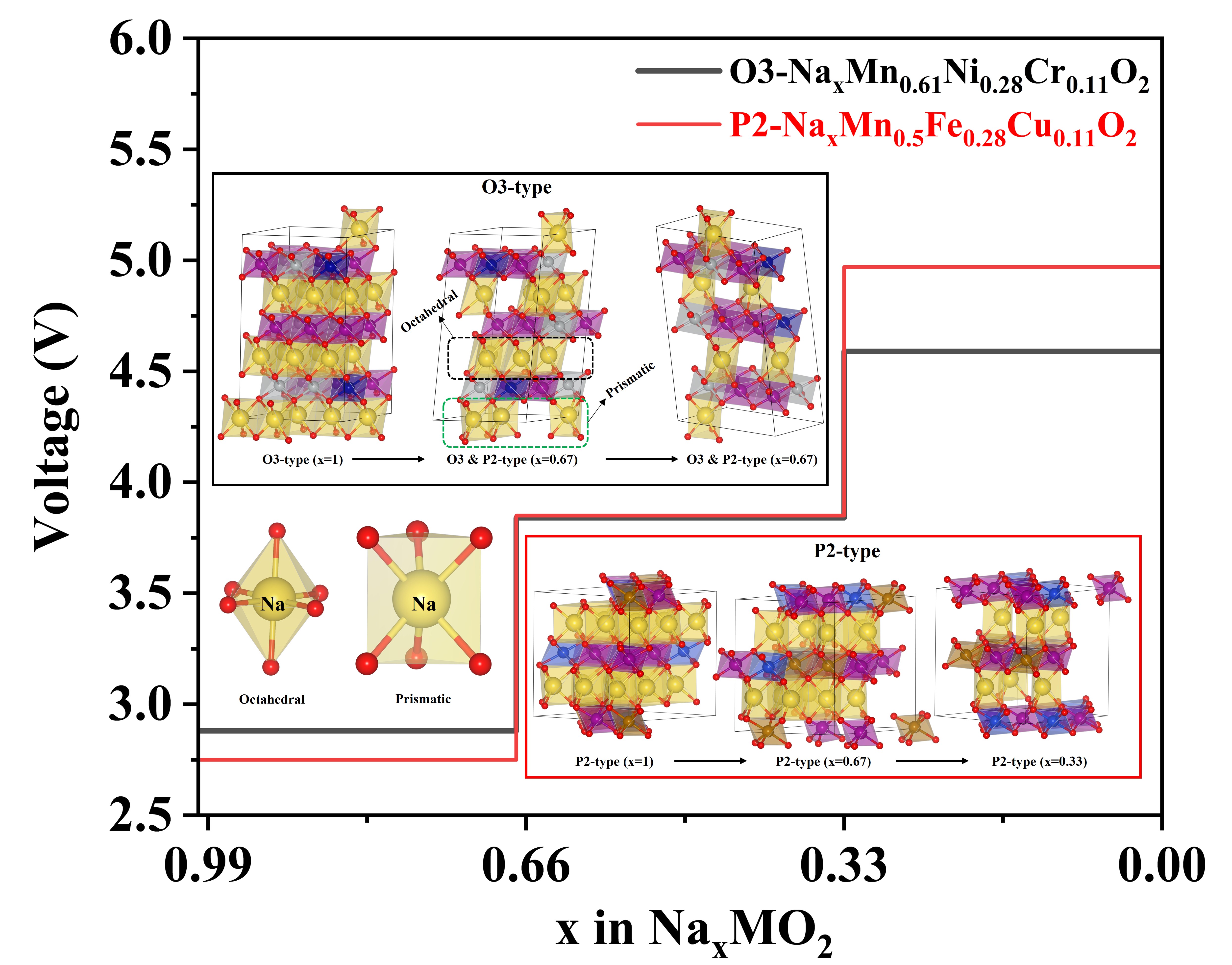}
    \caption{DFT calculated voltage profile of O3-Na$_{x}$Mn$_{0.61}$Ni$_{0.28}$Cr$_{0.11}$O$_{2}$ (black) and P2-Na$_{x}$Mn$_{0.5}$Fe$_{0.28}$Cu$_{0.22}$O$_{2}$ (red) during desodiation (x=1 to 0). Structural insets show coordination changes: O3-type undergoes a transition from octahedral to prismatic geometry, while P2-type retains prismatic coordination, indicating superior structural stability.}
    \label{fig:fig_8}
\end{figure}

Density functional theory (DFT) calculations were employed to evaluate the electrochemical properties of the newly designed cathode materials: O3-type Na$_{x}$Mn$_{0.61}$Ni$_{0.28}$Cr$_{0.11}$O$_{2}$ and P2-type Na$_{x}$Mn$_{0.5}$Fe$_{0.28}$Cu$_{0.22}$O$_{2}$, where x = 1, 0.67 and 0.33. The corresponding voltage profiles during desodiation (i.e., discharge from x = 1 to x=0) are presented in Figure \ref{fig:fig_8}. The calculated voltage range for the O3-type material spans from 2.88 to 4.59 V, while the P2-type composition exhibits a slightly wider range from 2.75 to 4.97 V.
The stepped features observed in the voltage profiles indicate the possibility of intermediate phase transitions during the desodiation process. The structural analysis (inset of Figure \ref{fig:fig_8}) reveals that the composition of type O3 undergoes a coordination transformation from octahedral (at x = 1) to mixed octahedral–prismatic (at x = 0.67) and finally to purely prismatic coordination at x=0.33, consistent with voltage steps and indicating progressive phase transformation. 

In contrast, the P2-type Na$_{x}$Mn$_{0.5}$Fe$_{0.28}$Cu$_{0.22}$O$_{2}$ maintains a stable prismatic Na coordination throughout the desodiation process (from x = 1 to x=0.33), suggesting superior structural robustness and phase stability (inset of Figure \ref{fig:fig_8}) during cycling. This invariance in the coordination environment underscores its potential for high-performance and long-life Na-ion battery applications. The incorporation of Cu into the P2-type material has been reported to enhance electronic conductivity and introduce additional high-voltage redox activity via the Cu$^{2+}$/Cu$^{3+}$ couple \cite{yang2021cu,wang2017copper,xu2023promising}, thereby supporting the observed broader voltage window. Importantly, the composition of the O3-type with intermediate Na content (i.e., Na$_{0.67}$Mn$_{0.61}$Ni$_{0.28}$Cr$_{0.11}$O$_{2}$), which exhibits a coexistence of octahedral and prismatic Na coordination, may offer an advantageous balance between structural flexibility and electrochemical performance, making it a promising candidate for further optimization. Cr in the O3-type composition stabilizes high-voltage redox activity through the Cr$^{3+}$/Cr$^{4+}$ couple\cite{cao2018reversible}, contributing to the increase in voltage beyond 4.0 V\cite{xu2023promising}.  Overall, these DFT insights validate the viability of the ML-predicted materials and provide a strong foundation for future experimental studies and the design of next-generation Na-ion battery cathodes.

\begin{figure}[htp]
    \centering
    \includegraphics[width=15cm]{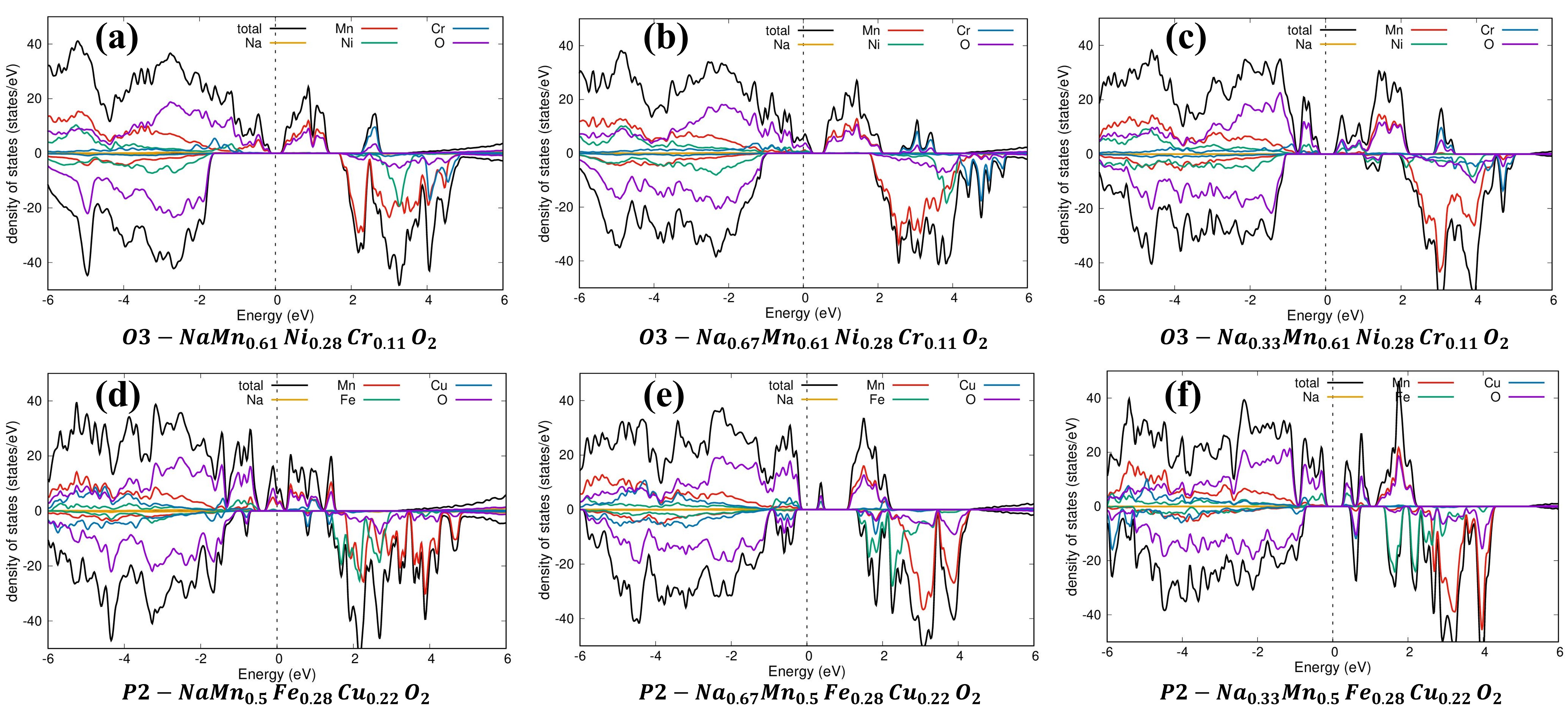}
    \caption{Spin-polarized total and projected density of states (DOS) for O3-type Na$_{x}$Mn$_{0.61}$Ni$_{0.28}$Cr$_{0.11}$O$_{2}$ at (a) x = 1, (b) x = 0.67, and (c) x = 0.33, and for P2-type Na$_{x}$Mn$_{0.5}$Fe$_{0.28}$Cu$_{0.22}$O$_{2}$ at (d) x = 1, (e) x = 0.67, and (f) x = 0.33. The evolution of the DOS with Na content reveals a narrowing band gap in the O3 system and metallic character in the P2 system, indicating improved electronic conductivity upon desodiation and the potential for high-rate capability in the P2-type material.}
    \label{fig:fig_9}
\end{figure} 

\begin{figure}[htp]
    \centering
    \includegraphics[width=15cm]{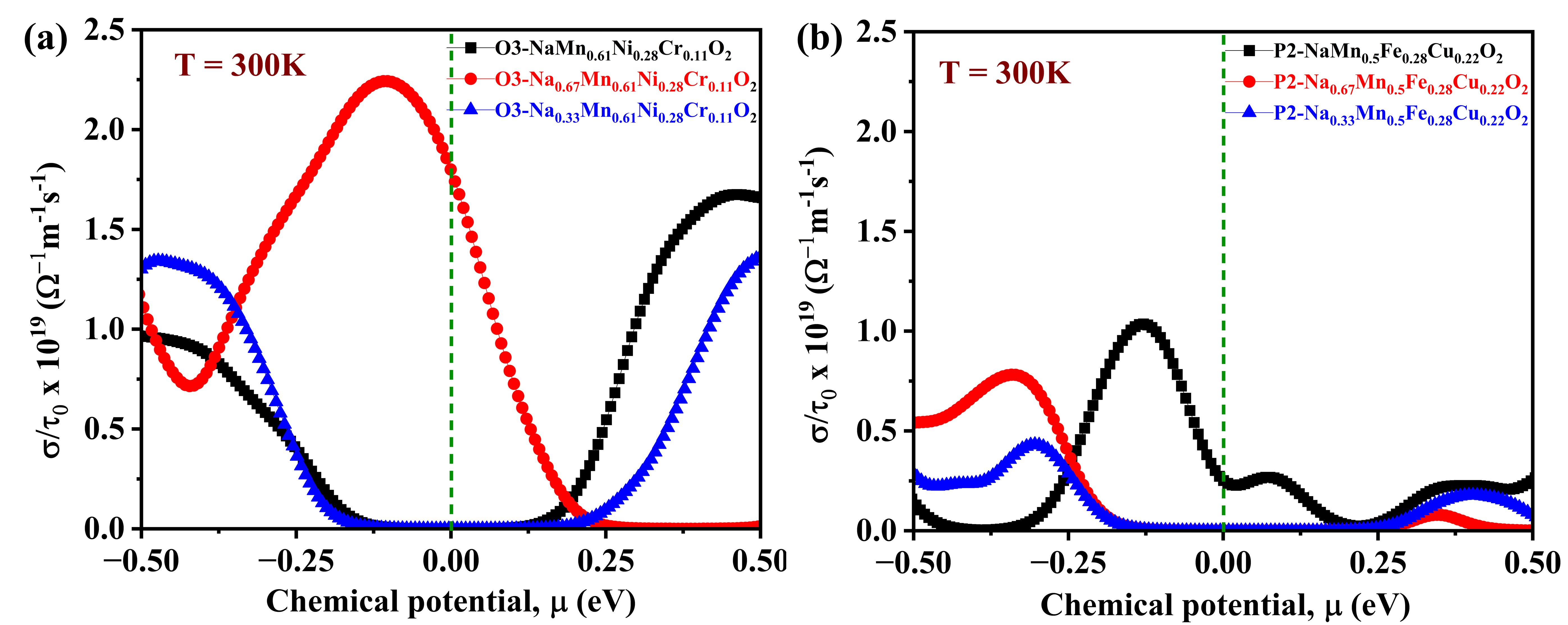}
    \caption{Electronic conductivity with respect to relaxation time ($\sigma/\tau$) as a function of chemical potential ($\mu$) at 300 K for (a) O3-type and (b) P2-type  compositions.}
    \label{fig:fig_10}
\end{figure}

To gain deeper insight into the electronic structure and conductivity of the ML-designed cathode materials, we computed the total and projected density of states (DOS) for both O3- and P2-type compositions with different sodium contents (x = 1, 0.67, and 0.33), as shown in Figure \ref{fig:fig_9}. O3-type Na$_{0.67}$Mn$_{0.61}$Ni$_{0.28}$Cr$_{0.11}$O$_{2}$ has higher DOS at Fermi level that can facilitate faster electron transport, as validated by the electrical conductivity data in Figure \ref{fig:fig_10}, can lead to improved power density and faster charging/discharging rates. For the O3-type Na$_{x}$Mn$_{0.61}$Ni$_{0.28}$Cr$_{0.11}$O$_{2}$ (Figures \ref{fig:fig_9} a–c), a narrowing of the band gap is observed upon desodiation, particularly from x = 1 to x = 0.33, indicating enhanced electronic conductivity at higher states of charge. The contributions from Ni and Cr 3d states dominate near the Fermi level, suggesting their active participation in the redox process. In contrast, the P2-type Na$_{x}$Mn$_{0.5}$Fe$_{0.28}$Cu$_{0.22}$O$_{2}$
(Figures \ref{fig:fig_9} d–f) exhibits a dramatic shift in the density of states (DOS) profile near the Fermi level, transitioning from metallic to semiconductor-like behavior. Notably, finite electronic states persist across all Na concentrations, with significant contributions at the Fermi level even at x = 1. Both Na$_{0.67}$Mn$_{0.5}$Fe$_{0.28}$Cu$_{0.22}$O$_{2}$ and Na$_{33}$Mn$_{0.5}$Fe$_{0.28}$Cu$_{0.22}$O$_{2}$ have large DOS near the Fermi level implying the availability of more number of states for electrons to occupy which reduces the energy barrier for electron excitation. The Cu and Fe 3d orbitals prominently contribute near the Fermi energy, indicating favorable electronic conductivity and robust charge-transport behavior during cycling. These electronic features support the observed voltage performance and suggest that the P2-type material may offer superior rate capability and electronic transport compared to the O3-type counterpart.

We further evaluate the electronic transport characteristics of materials designed with ML within the constant relaxation time approximation using BoltzTraP code. The transport coefficient, electronic conductivity normalized by relaxation time ($\sigma/\tau$) as a function of chemical potential ($\mu$) was calculated at 300 K, as shown in Figure \ref{fig:fig_10}. 
%For the O3-type Na$_{x}$Mn$_{0.61}$Ni$_{0.28}$Cr$_{0.11}$O$_{2}$ (Figure a), the conductivity is highest near the Fermi level for x = 0.67, suggesting enhanced charge carrier mobility at intermediate desodiation. In contrast, P2-type Na$_{x}$Mn$_{0.5}$Fe$_{0.28}$Cu$_{0.22}$O$_{2}$ Figure b) shows lower $\sigma/\tau$ values across all sodium concentrations, with minimal variation near the Fermi level. These findings imply that the O3-type material, especially at x = 0.67, may support faster electronic transport, potentially improving power performance in practical applications of Na-ion batteries.
The calculated electronic conductivity ($\sigma/\tau$) at 300 K for O3- and P2-type compositions reveals distinct trends across varying Na concentrations (x = 1, 0.67, and 0.33). For both structural types, $\sigma/\tau$ shows a non-monotonic dependence on the chemical potential ($\mu$), with a notable enhancement for intermediate Na content (x = 0.67). Specifically, O3-Na$_{x}$Mn$_{0.61}$Ni$_{0.28}$Cr$_{0.11}$O$_{2}$ exhibits the highest $\sigma/\tau$ near the Fermi level ($\mu \approx 0~\mathrm{eV}$), indicating favorable carrier mobility and improved electronic transport at this state of charge. Enhanced electrical conductivity leads to improved electrochemical behavior of the cathode \cite{jovanovic2018structural}. A similar trend is observed for P2-Na$_{x}$Mn$_{0.5}$Fe$_{0.28}$Cu$_{0.22}$O$_{2}$, albeit with a slightly lower magnitude. The enhanced $\sigma/\tau$ at x = 0.67 is correlated with mixed coordination environments and possible phase co-existence, which can facilitate a more effective charge delocalization. In contrast, compositions at x = 1 and 0.33 exhibit a reduced $\sigma/\tau$, likely due to more localized electronic states or increased structural distortion at high and low Na concentrations. These insights suggest that Na$_{0.67}$ stoichiometry in both O3 and P2 frameworks may offer an optimal balance between electronic conductivity and structural integrity, contributing to superior rate capability and overall battery performance.

\subsubsection{d-band center trends and electrochemical implications}
To better understand the origin of the electrochemical behavior, we evaluated the d-band center ($\epsilon_d$) positions of both O3- and P2-type materials at different Na concentrations. The d-band center is defined via~\cite{hammer1995electronic,bhattacharjee2016improved},
\begin{equation}
\varepsilon_d \;=\;\frac{\int_{-\infty}^{E_F} E\,D_d(E)\,dE}{\int_{-\infty}^{E_F} D_d(E)\,dE}
\label{d-band}
\end{equation}
Where $D_d(E)$ is the projected DOS onto transition-metal d states. $E_F$ is the Fermi energy.       

 An upward shift of the d-band center towards the Fermi energy level led to an enhanced Na$^{+}$ desorption behavior in the KNVO / NVO@C heterostructure (caused by the electric field of the interface at the heterojunction), which consequently accelerated Na$^{+}$ migration and facilitated diffusion of Na$^{+}$\cite{song2025tuning}. We observe a similar upshift in the d-band center with desodiation in the compositions of the O3 and P2 types for x = 1 to x = 0.67.  O3-type Na$_{x}$Mn$_{0.61}$Ni$_{0.28}$Cr$_{0.11}$O$_{2}$ exhibits a relatively shallow $\epsilon_d$, shifting from –1.305 eV at x = 1 to –1.246 eV at x = 0.33. In contrast, the P2-type Na$_{x}$Mn$_{0.5}$Fe$_{0.28}$Cu$_{0.22}$O$_{2}$ demonstrates deeper $\epsilon_d$ values, ranging from –1.862 eV to –1.965 eV over the same desodiation range. These trends are consistent with the theory of the d-band, where a higher (less negative) $\epsilon_d$ correlates with an increased hybridization between the transition metal (TM) d states and the O 2p orbitals, facilitating electronic conductivity and improving TM redox activity. For instance, O3-type materials, particularly at x = 0.67 ($\epsilon_d$ = –1.217 eV), present a favorable balance between redox potential and electronic transport, consistent with the enhanced $\sigma/\tau$ observed in Figure \ref{fig:fig_10}. This observation aligns with earlier studies [e.g. Ref\cite{grimaud2013double}, Ref\cite{seo2015calibrating}], which link shallow d-band centers to improved electrode kinetics and higher average voltages in layered oxide cathodes.
%
%\begin{table}[ht!]
%\centering
%\caption{d-band center (eV) of O3- and P2-type at different Na contents, indicating electronic structure evolution during desodiation.}
%    \begin{tabular}{lccr}
%    \headrow
%    \thead{O3-type}  & \thead{d-band center (eV)} \\ 
%    NaMn$_{0.61}$Ni$_{0.28}$Cr$_{0.11}$O$_2$  & -1.305 \\
%    Na$_{0.67}$Mn$_{0.61}$Ni$_{0.28}$Cr$_{0.11}$O$_2$  & -1.217 \\
%    Na$_{0.33}$Mn$_{0.61}$Ni$_{0.28}$Cr$_{0.11}$O$_2$  & -1.246 \\
%    \headrow
%    \thead{P2-type}  & \thead{d-band center (eV)} \\
%    \hline
%    NaMn$_{0.5}$Fe$_{0.28}$Cu$_{0.22}$O$_2$  & -1.862 \\
%    Na$_{0.67}$Mn$_{0.61}$Fe$_{0.28}$Cu$_{0.22}$O$_2$  & -1.678 \\
%    Na$_{0.33}$Mn$_{0.61}$Fe$_{0.28}$Cu$_{0.22}$O$_2$  & -1.965 \\    
%   \hline
%    \end{tabular}
%    \label{tab:table_dband}
%\end{table}
%
\begin{figure}[htp]
\centering
\includegraphics[width=15cm]{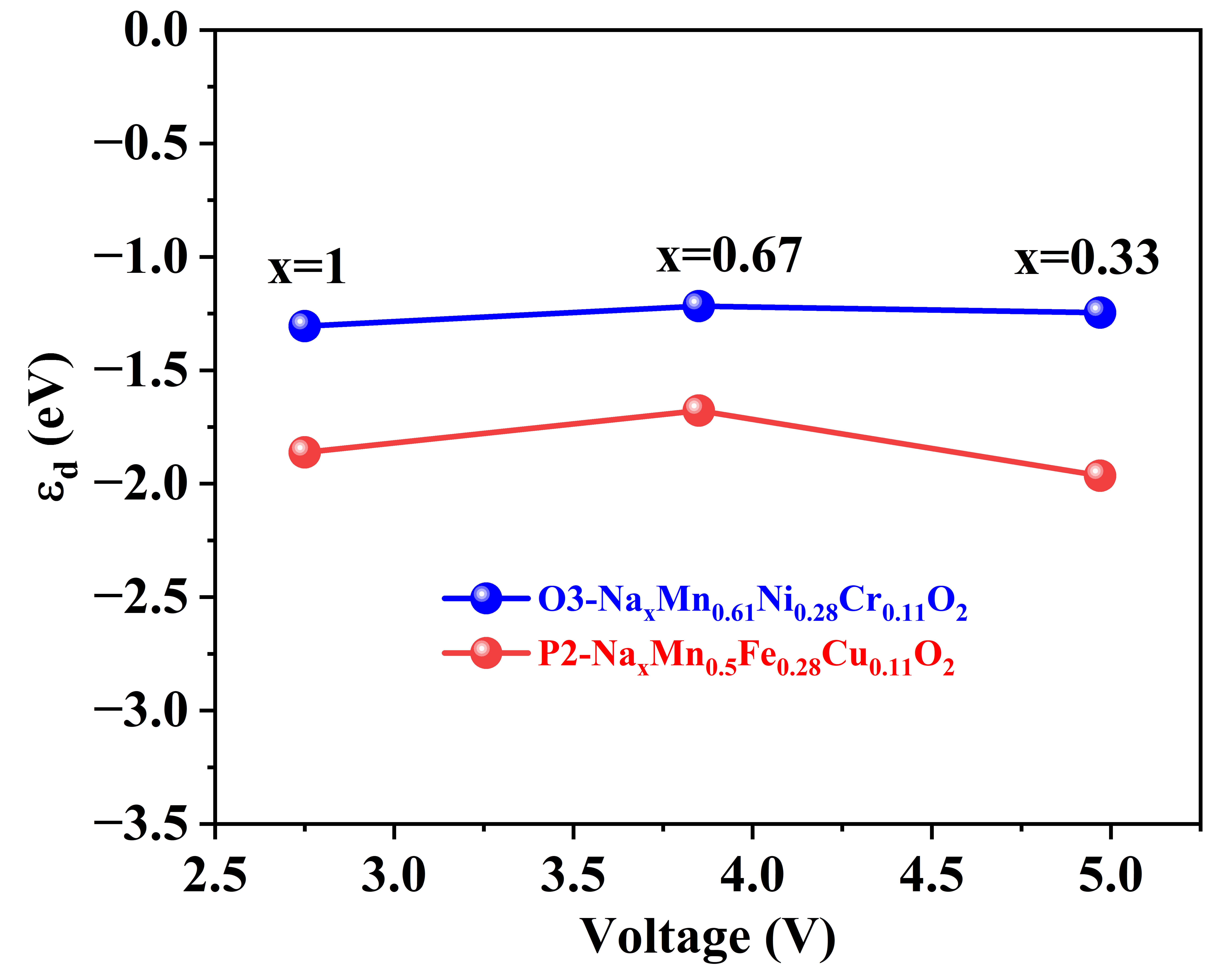}
\caption{$\varepsilon_d$ as a function of average cell voltage for O3–Na$_{x}$Mn$_{0.61}$Ni$_{0.28}$Cr$_{0.11}$O$_{2}$ and P2-Na$_{x}$Mn$_{0.5}$Fe$_{0.28}$Cu$_{0.22}$O$_{2}$}
\label{dc}
\end{figure}

Finally, Fig.\ref{dc} illustrates the evolution of the d-band center ($\varepsilon_d$) as a function of average cell voltage for the O3–Na$_{x}$Mn$_{0.61}$Ni$_{0.28}$Cr$_{0.11}$O$_{2}$ and P2-Na$_{x}$Mn$_{0.5}$Fe$_{0.28}$Cu$_{0.22}$O$_{2}$ cathode compositions. In both systems, $\varepsilon_d$ exhibits a non-monotonic dependence on sodium content, peaking near x=0.67, which corresponds to an intermediate state of desodiation. This shallowest $\varepsilon_d$ suggests enhanced hybridization between the transition-metal d states and oxygen 2p orbitals, facilitating improved electronic conductivity and redox activity. For the O3 phase, $\varepsilon_d$ remains consistently higher  than in the P2 counterpart, indicating stronger TM–O covalency and potentially more delocalized charge transport pathways. Notably, the observed peak in $\varepsilon_d$ coincides with the computed maxima in electronic conductivity ($\sigma/\tau$) mentioned earlier, reinforcing the notion that the d-band center serves as a meaningful descriptor linking electronic structure with electrochemical performance.

%\newpage
\section{Conclusions}
This study demonstrates the power of combining machine learning (ML) with first-principles simulations to accelerate the discovery and optimization of next-generation Na-ion battery cathodes. Our deep neural network (DNN) model, trained on Density Functional Theory (DFT) data, achieves a mean absolute error (MAE) of 0.24V, outperforming existing voltage prediction models with a high level of precision and a deviation of less than 5\%. To validate the predictive capacity of our model, we compared its voltage estimations with DFT-calculated values for two newly investigated cathode compositions: O3-Na$_{0.67}$Mn$_{0.61}$Ni$_{0.28}$Cr$_{0.11}$O$_{2}$ and P2-Na$_{0.67}$Mn$_{0.5}$Fe$_{0.28}$Cu$_{0.22}$O$_{2}$. The excellent agreement between the ML-predicted and DFT-computed voltages confirms the reliability of our approach in capturing the electrochemical behavior of Na-ion materials. By significantly reducing the reliance on time-consuming and resource-intensive experimental methods, our approach provides a systematic framework for identifying promising cathode materials with optimized electrochemical properties. This data-driven methodology enables rapid screening of novel compositions, guiding experimental efforts toward the most promising candidates.

The insights gained from this study contribute to the development of high-performance Na-ion batteries, offering a viable and sustainable alternative to Li-ion technology for future energy storage applications. Using machine learning-driven material discovery, we have helped pave the way for the rational design of next-generation battery materials, supporting advances in clean energy technologies and grid-scale storage solutions.

\section*{Acknowledgements}
This work was supported by the Korea Institute of Science and Technology (Grant number 2E31851), GKP (Global Knowledge Platform, Grant number 2V6760) project of the Ministry of Science, ICT and Future Planning. The authors thank Shubhayu Das of RWTH Aachen for useful discussions.

\section*{Conflicts of interest}
The authors declare that they have no conflict of interest.
\begin{suppinfo}
The supporting information file includes the definition of R$^2$ term, list of elemental characteristics used for the building of the DNN model. It also contains structural information on two specific compositions, the O3 and P2-type ordering (Figure S1). The formation energy of all the new O3 and P2-type ordering compositions is presented in Figure S2.
\end{suppinfo}
%\printendnotes
\bibliography{ref}
\begin{tocentry}
\includegraphics[width=8.4cm]{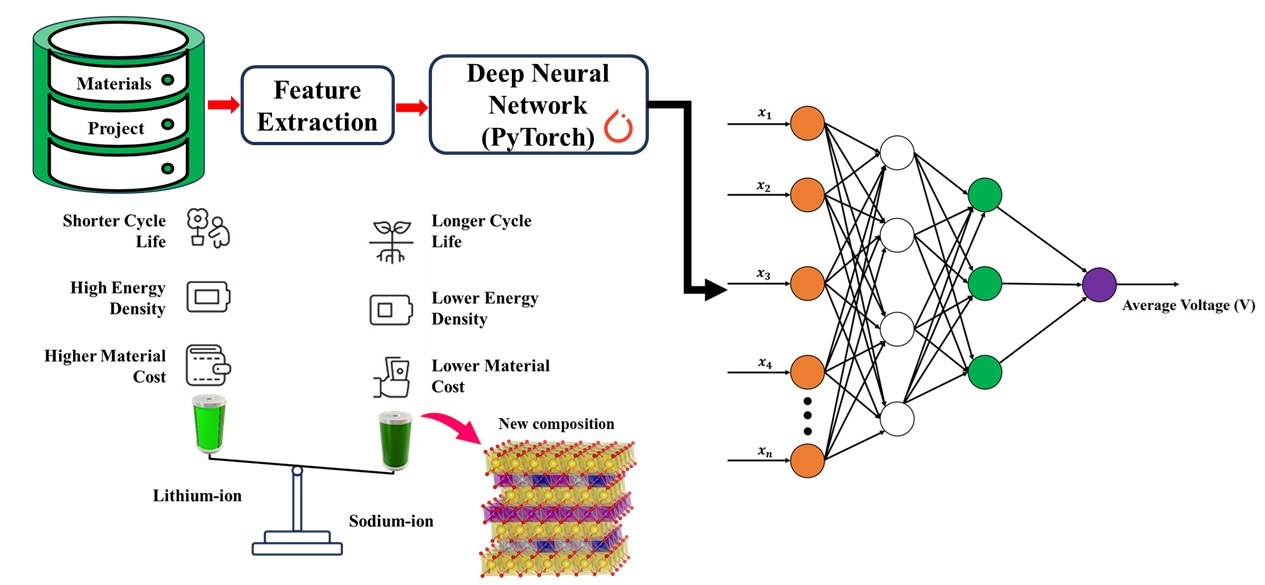}
\end{tocentry}
\end{document}